\documentclass[10pt]{article}
\usepackage{latexsym}
\usepackage{amsfonts}
\usepackage{indentfirst,bm,latexsym,amsmath,amsthm,amssymb}
\usepackage{graphicx}
\usepackage{subfigure}
\usepackage{color}
\usepackage{indentfirst}
\usepackage{titlesec}
\titleformat{\chapter}[display]
{\normalfont\Large\bfseries}{\thechapter}{11pt}{\Large}
\titleformat{\section}
{\normalfont\large\bfseries}{\thesection}{11pt}{\large}
\titlespacing*{\chapter}{0pt}{0pt}{15pt} 
\titlespacing*{\section}{0pt}{3.5ex plus 1ex minus .2ex}{2.3ex plus .2ex}
\usepackage{titletoc}
\usepackage[titletoc]{appendix}

\newcommand{\beqa}{\begin{eqnarray}}
\newcommand{\eeqa}{\end{eqnarray}}
\newcommand{\ba}{\begin{eqnarray*}}
\newcommand{\ea}{\end{eqnarray*}}

\theoremstyle{theorem}

\theoremstyle{remark}

\theoremstyle{theorem}

\theoremstyle{theorem}

\theoremstyle{theorem}

\renewcommand \theequation{\thesection.\arabic{equation}}
\begin{document}

\hoffset = -2.5truecm \voffset = -2.0truecm
\def\cdot{{\scriptstyle\,\bullet\,}}
\numberwithin{equation}{section}
\numberwithin{prop}{section}

\title{\bf Soliton solutions and their dynamics in reverse-space and reverse-space-time nonlocal discrete derivative nonlinear Schr\"odinger equations }
\author{ Gegenhasi\footnote{Corresponding author. E-mail: gegen@amss.ac.cn}, \quad Yue-Chen Jia
\\School of Mathematical Science, Inner Mongolia University,\\
No.235 West College Road, Hohhot, Inner Mongolia 010021,
 PR CHINA\\}

\maketitle
\begin{abstract}
\indent  In this paper, we introduce the reverse-space and reverse-space-time nonlocal discrete derivative nonlinear Schr\"odinger (DNLS) equations through the nonlocal symmetry reductions of the semi-discrete Gerdjikov-Ivanov equation. The muti-soliton solutions of two types of nonlocal discrete derivative nonlinear Schr\"odinger equations are derived by means of the Hirota bilinear method and reduction approach. We also investigate the dynamics of soliton solutions and reveal the rich soliton structures in the reverse-space and reverse-space-time nonlocal discrete DNLS equations. Our investigation shows that the solitons of these nonlocal equations often breathe and periodically collapse for some soliton parameters, but remain nonsingular for other range of parameters.

\end{abstract}

\section{Introduction}

Since Ablowitz and Musslimani proposed continuous and discrete reverse-space, reverse-time and reverse-space-time nonlocal nonlinear integrable equations by introducing new nonlocal symmetry reductions of the AKNS scattering problem and Ablowitz-Ladik scattering problem \cite{Abl1,Abl2,Abl3}, the nonlocal integrable equations have triggered renewed interest in integrable systems. A variety of mathematical methods such as inverse scattering methods \cite{Abl1,Abl2,Abl3,Abl4,Yang1}, Darboux transformation methods \cite{WYY,MSZ,Yang2}, Hirota's bilinear method and KP hierarchy reduction method \cite{FLAM,DLZ,XC,GP1,GP2,MZ} have been applied to study the nonlocal integrable equations. The nonlocal integrable equations possess some specific solution behaviors, such as finite-time solution blowup\cite{Abl1,Yang3}, the simultaneous existence of solitons and kinks\cite{JZ}, the simultaneous existence of bright and dark solitons\cite{Abl1,Abl4}, and distinctive multisoliton patterns\cite{Yang4}.

In \cite{Tsuchida}, the author proposed an integrable semi-discrete Gerdjikov-Ivanov equation
\begin{equation}\label{GI}
\left\{\begin{array}{l}
{i q_{n, t}+\left(q_{n+1}+q_{n-1}-2 q_{n}\right)-q_{n}\left(q_{n+1}+q_{n-1}\right)\left(r_{n+1}-r_{n}+q_{n} r_{n} r_{n+1}\right)=0}, \\
{i r_{n, t}-\left(r_{n+1}+r_{n-1}-2 r_{n}\right)+r_{n}\left(r_{n+1}+r_{n-1}\right)\left(q_{n-1}-q_{n}+r_{n} q_{n} q_{n-1}\right)=0},
\end{array}\right.
\end{equation}
where $q_n=q(n,t),r_n=r(n,t)$ are complex functions on $Z \times R$. The Miura map $u_{n}=q_n,v_n=r_{n+1}-r_{n}+q_{n} r_{n} r_{n+1}$ and another Miura map  $u_n=q_{n-1}-q_{n}+r_{n} q_{n} q_{n-1}, v_{n}=r_n$ connect the semi-discrete Gerdjikov-Ivanov equation
(\ref{GI}) with the coupled discrete nonlinear Schr\"odinger equation proposed by Ablowitz and Ladik
\begin{equation}\label{AL}
\left\{\begin{array}{l}
{i u_{n, t}+\left(u_{n+1}+u_{n-1}-2 u_{n}\right)-u_{n}v_n\left(u_{n+1}+u_{n-1}\right)=0}, \\
{i v_{n, t}-\left(v_{n+1}+v_{n-1}-2 v_{n}\right)+u_{n}v_n\left(v_{n+1}+v_{n-1}\right)=0}.
\end{array}\right.
\end{equation}
The semi-discrete Gerdjikov-Ivanov equation (\ref{GI}) has been solved by the inverse scattering method\cite{Tsuchida}. However, the Hirota bilinear formalism  of Eq.(\ref{GI})
has not been reported yet. In this paper, we present the bilinear form of the semi-discrete Gerdjikov-Ivanov equation (\ref{GI}) and obtain its one-, two- and three-soliton
solutions via Hirota bilinear method. It is known that the semi-discrete Gerdjikov-Ivanov equation (\ref{GI}) admits the local reduction of complex conjugation $r_n=\pm iq^{\ast}_{n-\frac{1}{2}}$. In this paper, we introduce two new nonlocal symmetry reductions $r_n=\sigma q^{\ast}_{-n},\sigma=\pm 1$ and $r_n=\sigma q_{-n}(-t),\sigma=\pm 1$ of the semi-discrete Gerdjikov-Ivanov equation (\ref{GI}), and obtain two nonlocal discrete DNLS equations:
\begin{equation}\label{ndNLS1}
iq_{n,t}+(q_{n-1}+q_{n+1}-2q_{n})+\sigma q_{n}(q_{n-1}+q_{n+1})(q^{*}_{-n}-q^{*}_{-n-1}-\sigma q^{*}_{-n}q_nq^{*}_{-n-1})=0,
\end{equation}
and
\begin{equation}\label{ndNLS2}
iq_{n,t}+(q_{n-1}+q_{n+1}-2q_{n})+\sigma q_{n}(q_{n-1}+q_{n+1})(q_{-n}(-t)-q_{-n-1}(-t)-\sigma q_{-n}(-t)q_nq_{-n-1}(-t))=0,
\end{equation}
respectively. We derive one-, two- and three-soliton solutions for reverse-space discrete DNLS equation (\ref{ndNLS1}) and reverse-space-time discrete DNLS equation (\ref{ndNLS2}), and study the dynamics of these soliton solutions.

The paper is organized as follows. In Section 2, we derive one-, two- and three-soliton solutions for the semi-discrete Gerdjikov-Ivanov equation (\ref{GI}) by applying the Hirota bilinear method. In Section 3, one-, two- and three-soliton solutions for the reverse-space discrete DNLS equation (\ref{ndNLS1}) are derived through the reduction approach and dynamics of these solitons are discussed.  In Section 4, we derive one-, two- and three-soliton solutions for the reverse-space-time discrete DNLS equation (\ref{ndNLS2}) via the reduction approach and investigate rich dynamics of soliton solutions. We end this paper with a conclusion and discussion in Section 5.

\section{Soliton solutions for the semi-discrete Gerdjikov-Ivanov equation (\ref{GI})}

In this section, we first bilinearise the semi-discrete Gerdjikov-Ivanov equation (\ref{GI}) and derive its one-, two- and three-soliton solutions via the Hirota bilinear method\cite{HB}.

Through the dependent variable transformations
\begin{equation}
q_{n}=\frac{g_{n}}{f_{n}}, r_{n}=-\frac{h_n}{s_{n}},
\end{equation}
Eq.(\ref{GI}) is transformed into the bilinear form
\begin{equation}\label{bl}
\left\{\begin{array}{l}
{iD_{t} f_{n} \cdot g_{n}=f_{n-1}g_{n+1}+f_{n+1}g_{n-1}-2 f_{n}g_{n}}, \\
{iD_{t} h_{n} \cdot s_{n}=h_{n+1}s_{n-1}+h_{n-1}s_{n+1}-2 h_{n}s_{n}}, \\
{g_{n}h_{n}-f_{n}s_{n}+f_{n-1} s_{n+1}=0}, \\
{g_{n} h_{n+1}+f_{n} s_{n+1}-f_{n+1} s_{n}=0},
\end{array}\right.
\end{equation}
where the bilinear operator $D_x^{m}D_t^{n}$  is defined by \cite{HB}
\begin{eqnarray*}
&& D_x^{m}D_t^{n}f\cdot g=\frac{\partial^m}{\partial
y^m}\frac{\partial^n}{\partial
s^n}f(x+y,t+s)g(x-y,t-s)|_{s=0,y=0}.
\end{eqnarray*}

According to Hirota bilinear method, in order to construct one-soliton solution, we expand the functions $g_{n}$, $f_{n}$ , $h_{n}$ and $s_{n}$ with a small parameter $\varepsilon$ as
\begin{equation}\label{expansion}
g_{n}=\varepsilon g^{(1)}_{n},\quad
h_{n}=\varepsilon h^{(1)}_{n},\quad
f_{n}=1+\varepsilon^{2} f^{(2)}_{n},\quad
s_{n}=1+\varepsilon^{2} s^{(2)}_{n}.
\end{equation}
By inserting expansions (\ref{expansion}) into bilinear equations (\ref{bl}), we obtain the coefficient of $\varepsilon^{1}$
\begin{equation}\label{first}
-i g^{(1)}_{n, t}=g^{(1)}_{n+1}+g^{(1)}_{n-1}-2 g^{(1)}_{n}, \quad i h^{(1)}_{n, t}=h^{(1)}_{n+1}+h^{(1)}_{n-1}-2 h^{(1)}_{n}.
\end{equation}
If we take the solution of linear differential-difference equations (\ref{first}) in the form
\begin{equation}\label{one}
g^{(1)}_{ n}=e^{\xi}, h^{(1)}_{n}=e^{\eta},
\end{equation}
with $\xi=kn+\omega t+\delta, \eta=ln+\rho t+\alpha$, then we yield the dispersion relations
\begin{equation}\label{1dispersion}
\omega=4i\sinh^2\frac{k}{2}, \quad \rho=-4i\sinh^2\frac{l}{2}.
\end{equation}
The coefficient of $\varepsilon^{2}$ gives
\begin{equation}\label{second}
g^{(1)}_{n} h^{(1)}_{n}-s^{(2)}_{n}-f^{(2)}_{n}+s^{(2)}_{n+1}+f^{(2)}_{n-1}=0,\quad g^{(1)}_{n} h^{(1)}_{n+1}+s^{(2)}_{n+1}+f^{(2)}_{n}-s^{(2)}_{n}-f^{(2)}_{n+1}=0.
\end{equation}
We obtain a solution of linear differential-difference equations (\ref{second}) in the exponential form
\begin{equation}\label{two}
f_{2,n}=A e^{\xi+\eta}, \quad s_{2,n}=B e^{\xi+\eta},
\end{equation} where
\begin{equation}\label{1coeff1}
A=\frac{e^{l}-1}{4 \sinh ^{2} \frac{k+l}{2}}, \quad B=\frac{e^{-k}-1}{4 \sinh ^{2} \frac{k+l}{2}}.
\end{equation}
It can be verified that the coefficients of $\varepsilon^{3},\varepsilon^{4}$ are automatically satisfied if we substitute (\ref{one}) and (\ref{two}) into them. Therefore, one-soliton solution of the semi-discrete Gerdjikov-Ivanov equation (\ref{GI}) is given by
\begin{equation}\label{1soliton}
q_{n}=\frac{e^{\xi}}{1+A e^{\xi+\eta}},\qquad r_{n}=-\frac{e^{\eta}}{1+B e^{\xi+\eta}},
\end{equation}
with $\xi=kn+(4i\sinh^2\frac{k}{2}) t+\delta,\eta=ln-(4i\sinh^2\frac{l}{2})t+\alpha, A=\frac{e^{l}-1}{4 \sinh ^{2} \frac{k+l}{2}}$ and  $B=\frac{e^{-k}-1}{4 \sinh ^{2} \frac{k+l}{2}}$. Here $k,l,\delta$ and $\alpha$ are arbitrary complex parameters.

For two-soliton solution, we take
\begin{equation}\label{expansion1}
g_{n}=\varepsilon g^{(1)}_{n}+\varepsilon^{3} g^{(3)}_{n},\quad h_{n}=\varepsilon h^{(1)}_{n}+\varepsilon^{3} h^{(3)}_{n},\quad f_{n}=1+\varepsilon^{2} f^{(2)}_{n}+\varepsilon^{4} f^{(4)}_{n},\quad s_{n}=1+\varepsilon^{2} s^{(2)}_{n}+\varepsilon^{4} s^{(4)}_{n}.
\end{equation}
When we insert expansions (\ref{expansion1}) into (\ref{bl}) and consider the coefficients of $\varepsilon$, we derive
\begin{equation*}
g^{(1)}_{n}=e^{\xi_{1}}+e^{\xi_{2}}, \ \ h^{(1)}_{n}=e^{\eta_{1}}+e^{\eta_{2}},
\end{equation*}
with $\xi_{j}=k_{j} n+\omega_{j} t+\delta_{j}$ , $\eta_{j}=l_{j} n+\rho_{j} t+\alpha_{j}$ for $j=1,2$,
and the dispersion relations
\begin{equation}\label{dispersion2}
\omega_{j}=4i \sinh ^{2}\frac{k_{j}}{2}, \quad \rho_{j}=-4i \sinh^{2} \frac{l_{j}}{2}, \quad  j=1,2.
\end{equation}
From the coefficient of $\varepsilon^{2}$, we derive
\begin{equation*}
f^{(2)}_{n}=e^{\xi_{1}+\eta_{1}+\alpha_{1,1}}+e^{\xi_{1}+\eta_{2}+\alpha_{1,2}}
+e^{\xi_{2}+\eta_{1}+\alpha_{2,1}}+e^{\xi_{2}+\eta_{2}+\alpha_{2,2}},
\end{equation*}
\begin{equation*}
s^{(2)}_{n}=e^{\xi_{1}+\eta_{1}+\delta_{1,1}}+e^{\xi_{1}+\eta_{2}+\delta_{1,2}}
+e^{\xi_{2}+\eta_{1}+\delta_{2,1}}+e^{\xi_{2}+\eta_{2}+\delta_{2,2}},
\end{equation*}
where
\begin{equation}\label{2coeff1}
e^{\alpha_{m,j}}=\frac{e^{l_{j}}-1}{4 \sinh ^{2} \frac{k_{m}+l_{j}}{2}},\quad e^{\delta_{m,j}}=\frac{e^{-k_{m}}-1}{4 \sinh ^{2} \frac{k_{m}+l_{j}}{2}},\ m,j=1,2.
\end{equation}
The coefficient of $\varepsilon^{3}$ gives
\begin{equation*}
g^{(3)}_{n}=\hat{A}_{1} e^{\xi_{1}+\xi_{2}+\eta_{2}}+\hat{A}_{2} e^{\xi_{1}+\xi_{2}+\eta_{2}},\quad h^{(3)}_{n}=\hat{B}_{1} e^{\xi_{1}+\eta_{1}+\eta_{2}}+\hat{B}_{2}, e^{\xi_{2}+\eta_{1}+\eta_{2}},
\end{equation*}
where
\begin{equation}\label{2coeff2}
\hat{A}_{m}=\left(e^{l_{m}-1}\right) \frac{\sinh ^{2} \frac{k_{1}-k_{2}}{2}}{4 \sinh ^{2} \frac{k_{1}+l_{m}}{2} \sinh ^{2} \frac{k_{2}+l_{m}}{2}},
\hat{B}_{m}=\left(e^{-k_{m}-1}\right) \frac{\sinh ^{2} \frac{l_{1}-l_{2}}{2}}{4 \sinh ^{2} \frac{k_{m}+l_{1}}{2} \sinh ^{2} \frac{k_{m}+l_{2}}{2}},\ m=1,2.
\end{equation}
From the coefficient of $\varepsilon^{4}$, we derive
\begin{equation*}
f^{(4)}_{n}=Me^{\xi_{1}+\xi_{2}+\eta_{1}+\eta_{2}},\quad s^{(4)}_{n}=Ne^{\xi_{1}+\xi_{2}+\eta_{1}+\eta_{2}},
\end{equation*}
where
\begin{equation}\label{2coeff4}
M=\frac{\left(e^{l_{1}}-1\right)\left(e^{l_{2}}-1\right) \sinh^{2} \frac{k_{1}- k_{2}}{2} \sinh^{2} \frac{l_{1}-l_{2}}{2}}{16 \sinh ^{2} \frac{k_{1}+l_{1}}{2} \sinh ^{2} \frac{k_{1}+l_{2}}{2}\sinh^{2}\frac{k_{2}+l_{1}}{2} \sinh ^{2} \frac{k_{2}+l_{2}}{2}},
N=\frac{\left(e^{-k_{1}}-1\right)\left(e^{-k_{2}}-1\right) \sinh^{2} \frac{k_{1}- k_{2}}{2} \sinh^{2} \frac{l_{1}-l_{2}}{2}}{16 \sinh ^{2} \frac{k_{1}+l_{1}}{2} \sinh ^{2} \frac{k_{1}+l_{2}}{2}\sinh^{2}\frac{k_{2}+l_{1}}{2} \sinh ^{2} \frac{k_{2}+l_{2}}{2}}.
\end{equation}
It can be verified the coefficients of $\varepsilon^{5},\varepsilon^{6},\varepsilon^{7},\varepsilon^{8}$ are automatically satisfied. Therefore, two-soliton solution of the semi-discrete Gerdjikov-Ivanov equation (\ref{GI}) is given by
\begin{equation}\label{2soliton1}
q_{n}=\frac{e^{\xi_{1}}+e^{\xi_{2}}
+\hat{A}_{1} e^{\xi_{1}+\xi_{2}+\eta_{1}}
+\hat{A}_{2}e^{\xi_{1}+\xi_{2}+\eta_{2}}}{1+
e^{\xi_{1}+\eta_{1}+\alpha_{1,1}}+e^{\xi_{1}+\eta_{2}+\alpha_{1,2}}
+e^{\xi_{2}+\eta_{1}+\alpha_{2,1}}+e^{\xi_{2}+\eta_{2}+\alpha_{2,2}}
+M e^{\xi_{1}+\xi_{2}+\eta_{1}+\eta_{2}}},
\end{equation}
\begin{equation}\label{2soliton2}
r_{n}=-\frac{e^{\eta_{1}}+e^{\eta_{2}}
+\hat{B}_{1} e^{\xi_{1}+\eta_{1}+\eta_{2}}
+\hat{B}_{2}e^{\xi_{2}+\eta_{1}+\eta_{2}}}{1+
e^{\xi_{1}+\eta_{1}+\delta_{1,1}}+e^{\xi_{1}+\eta_{2}+\delta_{1,2}}
+e^{\xi_{2}+\eta_{1}+\delta_{2,1}}+e^{\xi_{2}+\eta_{2}+\delta_{2,2}}
+N e^{\xi_{1}+\xi_{2}+\eta_{1}+\eta_{2}}},
\end{equation}
with  $\xi_{m}=k_{m}n+(4i\sinh^2\frac{k_m}{2})t+\delta_{m},\eta_{m}=l_{m}n-(4i\sinh^2\frac{l_m}{2})t+\alpha_{m}( m=1,2)$, and the coefficients $\alpha_{m,j},\delta_{m,j},A_{m},B_{m},M,N$ are given by (\ref{2coeff1})-(\ref{2coeff4}). Here $k_{m},l_{m},\delta_{m}$ and $\alpha_{m} (m=1,2)$ are arbitrary complex parameters.

For three-soliton solution, we take
\begin{equation}\label{expansion2}
\begin{aligned}
&g_{n}=\varepsilon g^{(1)}_{ n}+\varepsilon^{3} g^{(3)}_{n}+\varepsilon^{5} g^{(5)}_{n},\quad h_{n}=\varepsilon h^{(1)}_{n}+\varepsilon^{3} h^{(3)}_{n}+\varepsilon^{5} h^{(5)}_{n},\\
&f_{n}=1+\varepsilon^{2} f^{(2)}_{n}+\varepsilon^{4} f^{(4)}_{n}+\varepsilon^{6} f^{(6)}_{n},\quad s_{n}=1+\varepsilon^{2} s^{(2)}_{n}+\varepsilon^{4} s^{(4)}_{n}+\varepsilon^{6} s^{(6)}_{ n}.
\end{aligned}
\end{equation}
By substituting expansions (\ref{expansion2}) into bilinear equations (\ref{bl}) and considering the coefficients of $\varepsilon$, we derive
\begin{equation*}
g^{(1)}_{n}=e^{\xi_{1}}+e^{\xi_{2}}+e^{\xi_{3}}, \ \ h^{(1)}_{n}=e^{\eta_{1}}+e^{\eta_{2}}+e^{\eta_{3}},
\end{equation*}
with $\xi_{j}=k_{j} n+\omega_{j} t+\delta_{j}$ , $\eta_{j}=l_{j} n+\rho_{j} t+\alpha_{j}$ for $j=1,2,3$,
and the dispersion relations
\begin{equation}\label{dispersion3}
\omega_{j}=4i \sinh ^{2}\frac{k_{j}}{2}, \quad \rho_{j}=-4i \sinh^{2} \frac{l_{j}}{2},  \quad j=1,2,3.
\end{equation}
The coefficient of $\varepsilon^{2}$ gives
\begin{equation*}
f^{(2)}_{n}=\sum_{1 \leq m, j \leq 3} e^{\xi_{m}+\eta_{j}+\alpha_{m,j}},\quad s^{(2)}_{n}=\sum_{1 \leq m, j \leq 3} e^{\xi_{m}+\eta_{j}+\delta_{m,j}},
\end{equation*}
where
\begin{equation}\label{3coeff1}
e^{\alpha_{m,j}}=\frac{e^{l_{j}}-1}{4 \sinh ^{2} \frac{k_{m}+l_{j}}{2}},\quad e^{\delta_{m,j}}=\frac{e^{-k_{m}}-1}{4 \sinh ^{2} \frac{k_{m}+l_{j}}{2}},\ m,j=1,2,3.
\end{equation}
The coefficient of $\varepsilon^{3}$ gives
\begin{equation*}
g^{(3)}_{n}=\sum_{1 \leq m< j \leq 3} \sum_{1 \leq \mu \leq 3} A_{m,j,\mu} e^{\xi_{m}+\xi_{j}+\eta_{\mu}},\quad h^{(3)}_{n}=\sum_{1 \leq m< j \leq 3} \sum_{1 \leq \mu \leq 3}B_{\mu,m,j} e^{\xi_{\mu}+\eta_{m}+\eta_{j}},
\end{equation*}
where
\begin{equation}\label{3coeff2}
A_{m,j,\mu}=\frac{(e^{l_{\mu}-1})\sinh ^{2} \frac{k_{m}-k_{j}}{2}}{4 \sinh ^{2} \frac{k_{m}+l_{\mu}}{2} \sinh ^{2} \frac{k_{j}+l_{\mu}}{2}},
B_{\mu,m,j}=\frac{(e^{-k_{\mu}-1})\sinh ^{2} \frac{l_{m}-l_{j}}{2}}{4 \sinh ^{2} \frac{k_{\mu}+l_{m}}{2} \sinh ^{2} \frac{k_{\mu}+l_{j}}{2}},\ m,j,\mu\in\{1,2,3\},m<j.
\end{equation}
The coefficient of $\varepsilon^{4}$ gives
\begin{equation*}
f^{(4)}_{n}=\sum_{1 \leq m< j \leq 3} \sum_{1 \leq \mu< \nu \leq 3} M_{m,j,\mu,\nu} e^{\xi_{m}+\xi_{j}+\eta_{\mu}+\eta_{\nu}},\quad s^{(4)}_{n}=\sum_{1 \leq m< j \leq 3} \sum_{1 \leq \mu< \nu \leq 3} N_{m,j,\mu,\nu} e^{\xi_{m}+\xi_{j}+\eta_{\mu}+\eta_{\nu}},
\end{equation*}
where
\begin{equation}\label{3coeff4}
M_{m,j,\mu,\nu}=\frac{\left(e^{l_{\mu}}-1\right)\left(e^{l_{\nu}}-1\right) \sinh^{2} \frac{k_{m}- k_{j}}{2} \sinh^{2} \frac{l_{\mu}-l_{\nu}}{2}}{16 \sinh ^{2} \frac{k_{m}+l_{\mu}}{2} \sinh ^{2} \frac{k_{m}+l_{\nu}}{2}\sinh^{2}\frac{k_{j}+l_{\mu}}{2} \sinh ^{2} \frac{k_{j}+l_{\nu}}{2}},
\end{equation}
\begin{equation}\label{3coeff5}
N_{m,j,\mu,\nu}=\frac{\left(e^{-k_{m}}-1\right)\left(e^{-k_{j}}-1\right) \sinh^{2} \frac{k_{m}- k_{j}}{2} \sinh^{2} \frac{l_{\mu}-l_{\nu}}{2}}{16 \sinh ^{2} \frac{k_{m}+l_{\mu}}{2} \sinh ^{2} \frac{k_{m}+l_{\nu}}{2}\sinh^{2}\frac{k_{j}+l_{\mu}}{2} \sinh ^{2} \frac{k_{j}+l_{\nu}}{2}}.
\end{equation}
The coefficient of $\varepsilon^{5}$ gives
\begin{equation*}
g^{(5)}_{n}=\sum_{1 \leqslant m<j \leqslant 3}\tilde{A}_{m,j} e^{\xi_{1}+\xi_{2}+\xi_{3}+\eta_{m}+\eta_{j}},h^{(5)}_{n}=\sum_{1 \leqslant m<j \leqslant 3}\tilde{B}_{m,j} e^{\eta_{1}+\eta_{2}+\eta_{3}+\xi_{m}+\xi_{j}},
\end{equation*}
where
\begin{equation}\label{3coeff6}
\tilde{A}_{m,j}=\frac{\left(e^{l_{m}}-1\right)\left(e^{l_{j}}-1\right) \sinh ^{2} \frac{l_{m}-l_{j}}{2} \sinh ^{2} \frac{k_{1}-k_{2}}{2} \sinh ^{2} \frac{k_{1}-k_{3}}{2} \sinh ^{2} \frac{k_{2}-k_{3}}{2}}{16 \sinh ^{2} \frac{k_{1}+l_{m}}{2} \sinh ^{2} \frac{k_{1}+l_{j}}{2} \sinh ^{2} \frac{k_{2}+l_{m}}{2} \sinh ^{2} \frac{k_{2}+l_{j}}{2} \sinh ^{2} \frac{k_{3}+l_{m}}{2} \sinh ^{2} \frac{k_{3}+l_{j}}{2}},
\end{equation}
\begin{equation}\label{3coeff7}
\tilde{B}_{m,j}=\frac{\left(e^{-k_{m}}-1\right)\left(e^{-k_{j}}-1\right) \sinh ^{2} \frac{k_{m}-k_{j}}{2} \sinh ^{2} \frac{l_{1}-l_{2}}{2} \sinh ^{2} \frac{l_{1}-l_{3}}{2} \sinh ^{2} \frac{l_{2}-l_{3}}{2}}{16 \sinh ^{2} \frac{k_{m}+l_{1}}{2} \sinh ^{2} \frac{k_{j}+l_{1}}{2} \sinh ^{2} \frac{k_{m}+l_{2}}{2} \sinh ^{2} \frac{k_{j}+l_{2}}{2} \sinh ^{2} \frac{k_{m}+l_{3}}{2} \sinh ^{2} \frac{k_{j}+l_{3}}{2}}.
\end{equation}
The coefficient of $\varepsilon^{6}$ gives
\begin{equation*}
f^{(6)}_{n}=J e^{\xi_{1}+\xi_{2}+\xi_{3}+\eta_{1}+\eta_{2}+\eta_{3}},\quad s^{(6)}_{n}=K e^{\xi_{1}+\xi_{2}+\xi_{3}+\eta_{1}+\eta_{2}+\eta_{3}},
\end{equation*}
where
\begin{eqnarray}
&&J=\frac{\prod\limits_{p\in\{1,2,3\}}(e^{l_p}-1)][\prod\limits_{\substack{m,j\in\{1,2,3\}\\
m<j}}\sinh^{2} \frac{k_{m}-k_{j}}{2}\sinh^{2}\frac{l_{m}-l_{j}}{2}}{64\prod\limits_{p,\mu\in\{1,2,3\}}\frac{1}{\sinh^{2}\frac{k_{p}+l_{\mu}}{2}}},\label{3coeff8}\\
&& K=\frac{\prod\limits_{p\in\{1,2,3\}}(e^{-k_p}-1)][\prod\limits_{\substack{m,j\in\{1,2,3\}\\
m<j}}\sinh^{2} \frac{k_{m}-k_{j}}{2}\sinh^{2}\frac{l_{m}-l_{j}}{2}}{64\prod\limits_{p,\mu\in\{1,2,3\}}\frac{1}{\sinh^{2}\frac{k_{p}+l_{\mu}}{2}}}.\label{3coeff9}
\end{eqnarray}
\normalsize{It can be verified that the coefficients of $\varepsilon^{7},\varepsilon^{8},\varepsilon^{9},\varepsilon^{10},\varepsilon^{11},\varepsilon^{12}$ are automatically satisfied. Therefore, the semi-discrete Gerdjikov-Ivanov equation (\ref{GI}) has three-soliton solution in the form}
\small{\begin{eqnarray}
&&q_{n}=\frac{e^{\xi_{1}}+e^{\xi_{2}}+e^{\xi_{3}}
+\sum\limits_{1 \leq m< j \leq 3} \sum\limits_{1 \leq \mu \leq 3} A_{m,j,\mu} e^{\xi_{m}+\xi_{j}+\eta_{\mu}}
+\sum\limits_{1 \leqslant m<j \leqslant 3}\tilde{A}_{m,j} e^{\xi_{1}+\xi_{2}+\xi_{3}+\eta_{m}+\eta_{j}}}
{1+\sum\limits_{1 \leq m, j \leq 3} e^{\xi_{m}+\eta_{j}+\alpha_{m,j}}
+ \sum\limits_{1 \leq m< j \leq 3} \sum\limits_{1 \leq \mu< \nu \leq 3}M_{m,j,\mu,\nu} e^{\xi_{m}+\xi_{j}+\eta_{\mu}+\eta_{\nu}}
+J e^{\xi_{1}+\xi_{2}+\xi_{3}+\eta_{1}+\eta_{2}+\eta_{3}}},\label{3soliton1}\\
&&r_{n}=-\frac{e^{\eta_{1}}+e^{\eta_{2}}+e^{\eta_{3}}
+\sum\limits_{1 \leq m< j \leq 3} \sum\limits_{1 \leq \mu \leq 3}B_{\mu,m,j} e^{\xi_{\mu}+\eta_{m}+\eta_{j}}
+\sum\limits_{1 \leqslant m<j \leqslant 3}\tilde{B}_{m,j} e^{\eta_{1}+\eta_{2}+\eta_{3}+\xi_{m}+\xi_{j}}}
{1+\sum\limits_{1 \leq m, j \leq 3} e^{\xi_{m}+\eta_{j}+\delta_{m,j}}
+\sum\limits_{1 \leq m< j \leq 3} \sum\limits_{1 \leq \mu< \nu \leq 3} N_{m,j,\mu,\nu} e^{\xi_{m}+\xi_{j}+\eta_{\mu}+\eta_{\nu}}
+Ke^{\xi_{1}+\xi_{2}+\xi_{3}+\eta_{1}+\eta_{2}+\eta_{3}}},\label{3soliton2}
\end{eqnarray}}
\normalsize{with  $\xi_{j}=k_{j}n+(4i\sinh^2\frac{k_j}{2})t+\delta_{j},\eta_{j}=l_{j}n-(4i\sinh^2\frac{l_j}{2})t+\alpha_{j} (j=1,2,3)$ and the coefficients $\alpha_{m,j},\delta_{m,j},\tilde{A}_{m,j},$\\
$\tilde{B}_{m,j}, A_{m,j,\mu}, B_{s,i,j}, M_{i,j,s,t}, N_{i,j,s,t},J,K$ are given by (\ref{3coeff1}-\ref{3coeff9}).
Here $k_{i},l_{i},\delta_{i}$ and $\alpha_{i} (i=1,2,3)$ are arbitrary complex parameters.}

\section[$r_{n}=\sigma q_{-n}^{*}$]{Soliton solitons for the reverse-space nonlocal discrete DNLS equation (\ref{ndNLS1})}

    In this section, we derive one-, two-, three-soliton solutions for the reverse-space DNLS equation (\ref{ndNLS1}) by
 finding the constraint conditions on the parameters of one-, two-, three-soliton solutions of the semi-discrete Gerdjikov-Ivanov equation (\ref{GI}) to satisfy the the reduction formula $r_{n}= \sigma q_{-n}^{*}$.

\subsection{One-soliton solutions}
From one-soliton solution (\ref{1soliton}) and reduction formula $r_{n}= \sigma q_{-n}^{*}$, we have
\begin{equation} \label{1reduction1}
-\frac{e^{l n+\rho t+ \alpha }}{1+B e^{(k+l) n+(\omega+\rho) t+\delta+\alpha}}=\frac{\sigma e^{-k^{*} n+\omega^{*} z+\delta^{*}}}{1+A^{*} e^{-\left(k^{*}+l^{*}\right) n+\left(\omega^{*}+\rho^{*}\right)t+\delta^{*}+\alpha^{*}} },
\end{equation}
which yields the constraint conditions on four free paramaters $k,l,\delta,\alpha:$
\begin{equation}\label{consitraint1}
\begin{aligned}
&(1)\ l=-k^{*}, \qquad (2)\ \rho=\omega^{*}, \qquad (3)\ e^{\alpha}=-\sigma e^{\delta^{*}}, \qquad (4)\ B=A^{*}, \\
&(5)\ k+l=-\left(k^{*}+l^{*}\right), \qquad (6) \ \rho+\omega=\omega^{*}+\rho^{*},  \qquad (7) \ e^{\delta+\alpha}=e^{\delta^{*}+\alpha^{*}}.
\end{aligned}
\end{equation}
Utilizing the dispersion relation (\ref{1dispersion}) and (\ref{1coeff1}), Eq.(\ref{consitraint1}) can be reduced to the following two constraints
\begin{equation}
(1) \ l=-k^{*},\quad (2) \ e^{\alpha}=-\sigma e^{\delta^{*}}.
\end{equation}
Therefore, the reverse-space discrete DNLS equation (\ref{ndNLS1}) has the following form of one soliton solution
\begin{equation}\label{n1soliton1}
q_{n}=\frac{e^{kn+(4i\sinh^2\frac{k}{2})t+\delta}}{1-A\sigma e^{(k-k^{*})n+4i(\sinh^2\frac{k}{2}-\sinh^2\frac{k^{*}}{2})t+(\delta+\delta^{*})}},
\end{equation}
where $A=\frac{e^{-k^{*}}-1}{4 \sinh^{2} \frac{k-k^{*}}{2}}$ and $k,\delta$ are arbitrary complex parameters.

By letting $k=a+bi,\delta=c+di,A=L+Mi$, we obtain
\begin{equation}\label{module1}
|q_{n}|^{2}=\frac{e^{2an}}{e^{-2R}+e^{2R}(L^2+M^2)-2\sigma\sqrt{L^2+M^2}\cos(2bn+\gamma)},
\end{equation}
where $R=c-2\sin(b)\sinh(a)t$ and $\gamma$ is determined by $\sin(\gamma)=\frac{M}{\sqrt{L^2+M^2}},\cos(\gamma)=\frac{L}{\sqrt{L^2+M^2}}.$
In the spaecial case $a=0$, (\ref{module1}) becomes
\begin{equation}\label{module11}
|q_{n}|^{2}=\frac{1}{e^{-2c}+e^{2c}(L^2+M^2)-2\sigma\sqrt{L^2+M^2}\cos(2bn+\gamma)},
\end{equation}
which is a spatial periodical solution with the period $\frac{\pi}{b}$. By taking parameters as $k=2i,\delta=3+4i,\sigma=-1,$ the spatial periodical solution (\ref{module11}) is illustrated in $(a)$ of Fig.1.

If $a\neq0,$  then one-soliton solution (\ref{n1soliton1}) would breathe and periodically collapse in $n$ at time $t=\frac{c+\frac{\ln(L^2+M^2)}{4}}{2\sin(b)\sinh(a)}$ and its amplitude $|q_n|$ changes as
\begin{equation}\label{module12}
|q_{n}|^{2}=\frac{\sqrt{L^2+M^2}e^{2an}}{2(1-\sigma\cos(2bn+\gamma))}.
\end{equation}
When $b\neq0,$ this soliton periodically collapses in $n$ with period $\frac{\pi}{b}$ and its amplitude grows or decays exponentially (depending on the sign of $a$), which are shown in $(a)$ and $(b)$ of Fig.2 by choosing the parameters as
\begin{equation*}
\begin{aligned}
&k=-0.3-0.7i, \delta=1+\pi i, \sigma=-1, \\
\end{aligned}
\end{equation*}
and
\begin{equation*}
\begin{aligned}
& k=0.4+0.9i,  \delta=1+\pi i, \sigma=-1,
\end{aligned}
\end{equation*}
respectively.

We obtain another type of one-soliton solution for the reverse-space discrete DNLS equation (\ref{ndNLS1}) by the cross multiplication reduction. Applying the cross multiplication on Eq.\eqref{1reduction1}, we obtain
\begin{equation}
-e^{ln+\rho t+\alpha}(1+A^{*}e^{-(k^{*}+l^{*})n+(\omega^{*}+\rho^{*})+\delta^{*}+\alpha^{*}})=\sigma e^{-k^{*}n+\omega^{*}t+\delta}(1+Be^{(k+l)n+(\omega+\rho)t+\delta+\alpha}),
\end{equation}
from which we derive the conditions
\begin{equation}\label{key3}
\begin{aligned}
&(1)\ k=k^{*},l=l^{*}  \\
&(2)\ e^{\delta+\delta^{*}}=-\frac{1}{\sigma B},
e^{\alpha+\alpha^{*}}=-\frac{\sigma}{A^{*}},
\end{aligned}
\end{equation}
in which $A=\frac{e^{l}-1}{ 4 \sinh^{2} \frac{k+l}{2}}$ and $B=\frac{e^{-k}-1}{ 4 \sinh^{2} \frac{k+l}{2}}$. Setting $\delta=a+bi$,$\alpha=c+di$, then
according to the Eq.{\eqref{key3}}, we obtain
\begin{equation}
\begin{aligned}
&(1) \ e^{a}=\sqrt{\frac{1}{-\sigma B}}, \\
&(2) \ e^{c}=\sqrt{\frac{1}{-\sigma A}},
\end{aligned}
\end{equation}
where $a,b,c,d,k,l$ are real.

Therefore, another type of one soliton solution for Eq.(\ref{ndNLS1}) is given by
\begin{equation}\label{key2}
	q(n,t)=\frac{ e^{kn+4i \sinh^{2} \frac{k}{2} t+bi}}{\sqrt{-\sigma B}(1+\sqrt{\frac{A}{B}}e^{(k+l)n+4i(\sinh^{2}\frac{k}{2}t-\sinh^{2}\frac{l}{2})t+(b+d)i})},
\end{equation}
where $b,d,k,l$ are free real parameters. The corresponding $|q_n|^{2}$ is
\begin{equation}\label{module2}
|q_{n}|^{2}=\frac{e^{2kn+2a}}{1+A^2e^{2(k+l)n+2(a+c)}+2A\cos(R)e^{(k+l)n+(a+c)}},
\end{equation}
where $R=4(\sinh^{2}\frac{k}{2}-\sinh^{2}\frac{l}{2})t+(b+d)$. From (\ref{module2}), we derive one-soliton solution (\ref{key2}) breathes and periodically collapses in time at position $n=\frac{\ln \frac {A}{B}}{2(k+l)}$, in which the condition $\frac{\ln \frac {A}{B}}{2(k+l)}\in Z$ should be satisfied. The period of this collapse is $\frac{\pi}{2(\sinh^{2}\frac{k}{2}-\sinh^{2}\frac{l}{2})}$.

The graph of one soliton solution (\ref{key2}) is depicted in $(b)$ of Fig.1 by taking the parameters:
\begin{equation*}
\sigma=-1,k=\ln(1-e^{-0.3}), l=0.3, b=1, d=1.
\end{equation*}

\begin{figure}[!htb]
\centering
\subfigure[]{
\begin{minipage}[t]{0.32\linewidth}
\centering
\includegraphics[width=2in]{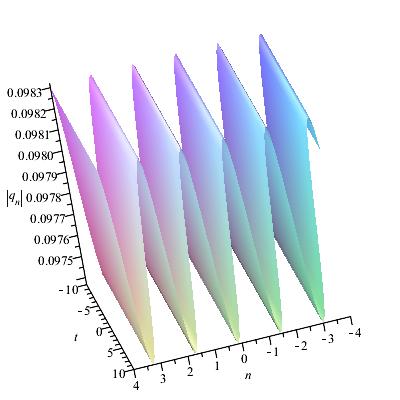}
\end{minipage}%
}%
\subfigure[]{
\begin{minipage}[t]{0.32\linewidth}
\centering
\includegraphics[width=2in]{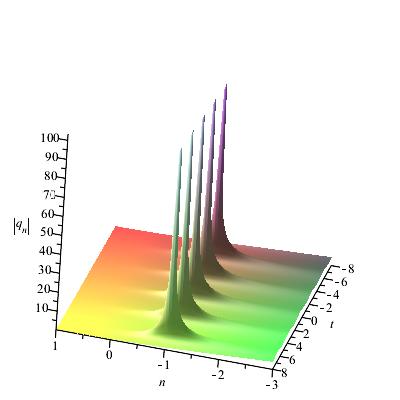}
\end{minipage}%
}%
\centering
\caption{One-soliton solution for Eq.(\ref{ndNLS1}): (a) Nonsingular spatial periodic solution, (b) solution breathing and periodically collapsing in time.}
\end{figure}

\begin{figure}[!htb]
\centering
\subfigure[]{
\begin{minipage}[t]{0.32\linewidth}
\centering
\includegraphics[width=2in]{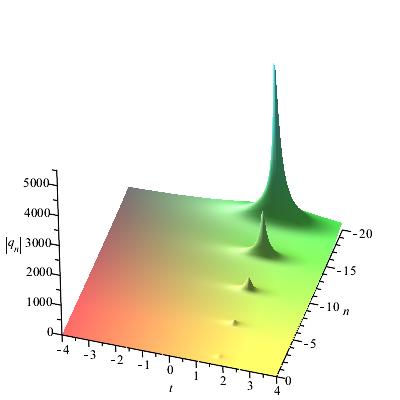}
\end{minipage}%
}%
\subfigure[]{
\begin{minipage}[t]{0.32\linewidth}
\centering
\includegraphics[width=2in]{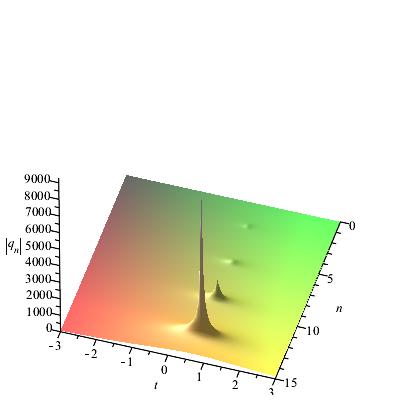}
\end{minipage}%
}%
\centering
\caption{One-soliton solution periodically collapsing in space: (a)Solution with exponentially growing amplitude, (b)Solution with exponentially decaying amplitude.}
\end{figure}

\subsection{Two-solitons}
From the two-soliton solution (\ref{2soliton1}-\ref{2soliton2}) and reduction formula $r_{n}= \sigma q_{-n}^{*}$, we have
\begin{equation}\label{reduc}
\begin{aligned}
-\frac{e^{\eta_{1}}+e^{\eta_{2}}
+\hat{B}_{1} e^{\xi_{1}+\eta_{1}+\eta_{2}}
+\hat{B}_{2}e^{\xi_{2}+\eta_{1}+\eta_{2}}}{1+
e^{\xi_{1}+\eta_{1}+\delta_{1,1}}+e^{\xi_{1}+\eta_{2}+\delta_{1,2}}
+e^{\xi_{2}+\eta_{1}+\delta_{2,1}}+e^{\xi_{2}+\eta_{2}+\delta_{2,2}}
+N e^{\xi_{1}+\xi_{2}+\eta_{1}+\eta_{2}}}
= \\
 \sigma
\frac{e^{\bar{\xi}_{1}^{*}}+e^{\bar{\xi}_{2}^{*}}
+\hat{A}_{1}^{*} e^{\bar{\xi}_{1}^{*}+\bar{\xi}_{2}^{*}+\bar{\eta}_{1}^{*}}
+\hat{A}_{2}^{*}e^{\bar{\xi}_{1}^{*}+\bar{\xi}_{2}^{*}+\bar{\eta}_{2}^{*}}}
{1+e^{\bar{\xi}_{1}^{*}+\bar{\eta_{1}}^{*}+\alpha_{1,1}^{*}}+e^{\bar{\xi}_{1}^{*}+\bar{\eta}_{2}^{*}+\alpha_{1,2}^{*}}
+e^{\bar{\xi}_{2}^{*}+\bar{\eta}_{1}^{*}+\alpha_{2,1}^{*}}+e^{\bar{\xi}_{2}^{*}+\bar{\eta}_{2}^{*}+\alpha_{2,2}^{*}}
+M e^{\bar{\xi}_{1}^{*}+\bar{\xi}_{2}^{*}+\bar{\eta}_{1}^{*}+\bar{\eta}_{2}^{*}}},
\end{aligned}
\end{equation}where $\bar{\xi}_{j}=-k_jn+\omega_j t+\delta_j,\bar{\eta}_{j}=-l_jn+\rho_jt+\alpha_j (j=1,2).$
 Eq.(\ref{reduc}) yields the constraint conditions on the eight paramaters $k_j,l_j,\delta_j,\alpha_j (j=1,2)$:
\begin{equation}
\begin{aligned}
&(1) \ l_{j}=-k_{j}^{*}, \ j=1,2,\quad  (2) a_{j}=\omega_{j}^{*},\ j=1,2, \quad (3) e^{\alpha_{j}}=- \sigma e^{\delta^{*}_{j}},\ j=1,2, \quad (4) \hat{B}_{j}=\hat{A}_{j}^{*},\ j=1,2, \\
&(5) \  k_{1}+l_{1}+l_{2}=-\left(k_{1}^{*}+k_{2}^{*}+l_{1}^{*}\right),\quad k_{2}+l_{1}+l_{2}=-\left(k_{1}^{*}+k_{2}^{*}+l_{2}^{*}\right), \quad (6) e^{\alpha^{*}_{m,j}}=e^{\delta_{j,m}},\ m,j=1,2 ,\\
&(7) \ \omega_{1}+\rho_{1}+\rho_{2}=-\left(\omega_{1}^{*}+\omega_{2}^{*}+\rho_{1}^{*},\right),\quad \omega_{2}+\rho_{1}+\rho_{2}=-\left(\omega_{1}^{*}+\omega_{2}^{*}+\rho_{2}^{*}\right), \quad  (8) \  N=M^{*}.\label{2constraints1}
\end{aligned}
\end{equation}
Utilizing the dispersion relations (\ref{dispersion2}) and Eqs.(\ref{2coeff1}-\ref{2coeff4}), Eq.(\ref{2constraints1}) can be reduced to the following four conditions
\begin{equation}\label{2constraint21}
(1) \ l_{j}=-k_{j}^{*}, \quad (2) \ e^{\alpha_{j}}=- \sigma e^{\delta_{j}^{*}}, \quad j=1,2.
\end{equation}
Therefore, the two-soliton solution for the reverse-space discrete DNLS equation (\ref{ndNLS1}) is given by (\ref{2soliton1}) with constraints of parameters (\ref{2constraint21}). The graph of this two-soliton solution is depicted in Fig.3 and Fig.4 by taking the parameters as
\begin{equation*}
\begin{aligned}
& k_{1}=0.2i, k_{2}=0.8i, \delta_{1}=1+2i,  \delta_{2}=i, \sigma=-1,
\end{aligned}
\end{equation*}
and
\begin{equation*}
\begin{aligned}
&(a)\quad k_{1}=0.3+0.6i, k_{2}=-0.4-0.9i, \delta_{1}=0,  \delta_{2}=0, \sigma=1,\\
&(b)\quad k_{1}=0.2+0.4i, k_{2}=-0.2-0.4i, \delta_{1}=0,  \delta_{2}=0, \sigma=1,
\end{aligned}
\end{equation*}
respectively.

\begin{figure}[!htb]
\centering
\subfigure[]{
\begin{minipage}[t]{0.32\linewidth}
\centering
\includegraphics[width=2in]{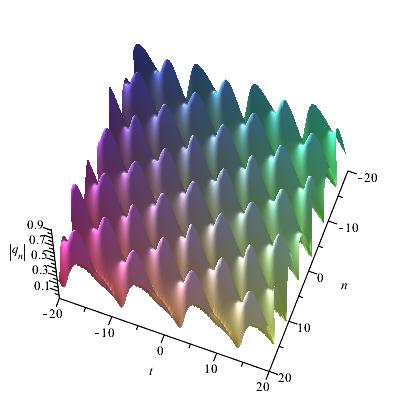}
\end{minipage}%
}%
\subfigure[]{
\begin{minipage}[t]{0.32\linewidth}
\centering
\includegraphics[width=2in]{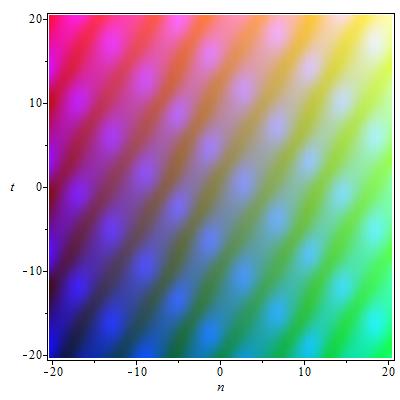}
\end{minipage}%
}%
\centering
\caption{Two-soliton solution for Eq.(\ref{ndNLS1}): (a)Nonsingular periodic two-soliton, (b)The density profiles of (a). }
\end{figure}

\begin{figure}[!htb]
\centering
\subfigure[]{
\begin{minipage}[t]{0.32\linewidth}
\centering
\includegraphics[width=2in]{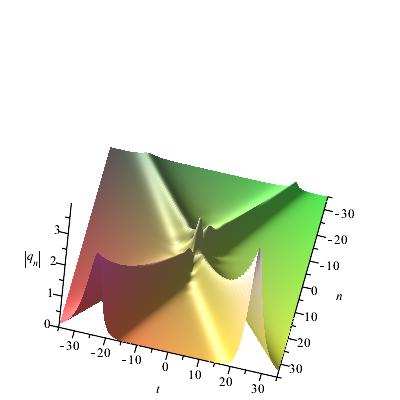}
\end{minipage}%
}%
\subfigure[]{
\begin{minipage}[t]{0.32\linewidth}
\centering
\includegraphics[width=2in]{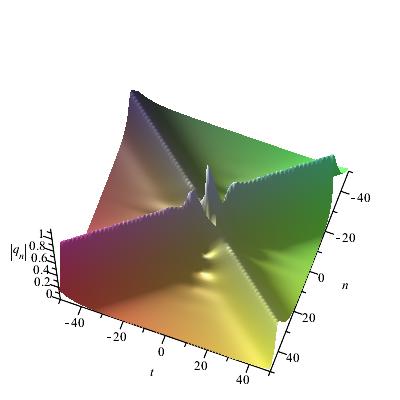}
\end{minipage}%
}%
\centering
\caption{Two-soliton solution for Eq.(\ref{ndNLS1}): (a)Two nonsingular solitons with changing amplitude moving in opposite directions, (b)Elastic collision of two soliton. }
\end{figure}

We derive another type of two-soliton solution for the reverse-space discrete DNLS equation (\ref{ndNLS1}) via the cross multiplication reduction. Applying the cross multiplication on {\eqref{reduc}}. we obtain the following constraints on eight paramaters $k_j,l_j,\delta_j,\alpha_j (j=1,2)$:
\begin{equation}\label{1key}
\begin{aligned}
&(1) \ k_{j}=k_{j}^{*}, l_{j}=l_{j}^{*} (j=1,2), \quad (2)\ e^{\delta_{1}+\delta_{1}^{*}}=-\frac{\hat{B}_{2}}{\sigma N},\\
&(3) \ e^{\delta_{2}+\delta_{2}^{*}}=-\frac{\hat{B}_{1}}{\sigma N},\quad (4) \ e^{\alpha_{1}+\alpha_{1}^{*}}=-\frac{\sigma \hat{A}_{2}^{*}}{M^{*}},(5) \ e^{\alpha_{2}+\alpha_{2}^{*}}=-\frac{\sigma \hat{A}_{1}^{*}}{M^{*}}.  \\
\end{aligned}
\end{equation}
We suppose $\delta_{j}=a_{j}+b_{j}i$,$\alpha_{j}=x_{j}+y_{j}i(j=1,2)$, where $a_{j},b_{j},x_{j},y_{j}(j=1,2)$  are real. According to {\eqref{1key}}, we obtain
\begin{equation}\label{2key}
	\begin{aligned}
	& \ (1)e^{a_{1}}=2 \sqrt{\frac{\sinh^{2} \frac{k_{1}+l_{1}}{2} \sinh^{2} \frac{k_{1}+l_{2}}{2}}{\sigma  (1-e^{-k_{1}})\sinh^{2}\frac{k_{1}-k_{2}}{2}}},
	& \ (2)e^{a_{2}}=2 \sqrt{ \frac{\sinh^{2} \frac{k_{2}+l_{1}}{2} \sinh^{2} \frac{k_{2}+l_{2}}{2}}{\sigma  (1-e^{-k_{2}})\sinh^{2} \frac{k_{1}-k_{2}}{2}}},   \\
	& \ (3) e^{x_{1}}=2 \sqrt{\frac{\sinh^{2} \frac{k_{1}+l_{1}}{2} \sinh^{2} \frac{k_{2}+l_{1}}{2}}{\sigma (1-e^{l_{1}})\sinh^{2} \frac{l_{1}-l_{2}}{2}}},
	& \ (4) e^{x_{2}}=2 \sqrt{\frac{\sinh^{2} \frac{k_{1}+l_{2}}{2} \sinh^{2} \frac{k_{2}+l_{2}}{2}}{\sigma (1-e^{l_{2}})\sinh^{2} \frac{l_{1}-l_{2}}{2}}}.
	\end{aligned}
\end{equation}
Therefore, another type of two-soliton solution for the reverse-space discrete DNLS equation (\ref{ndNLS1}) is given by (\ref{2soliton1}) with constraints of parameters (\ref{2key}).
We illustrate this two-soliton in Fig.5 by taking
\begin{equation*}
\begin{aligned}
&k_{1}=0.3, k_{2}=0.8, l_{1}=0.3, l_{2}=0.8, b_{1}=0, b_{2}=0, y_{1}=0, y_{2}=0, \sigma=1.
\end{aligned}
\end{equation*}

\begin{figure}[!htb]
\centering
\subfigure[]{
\begin{minipage}[t]{0.29\linewidth}
\centering
\includegraphics[width=1.8in]{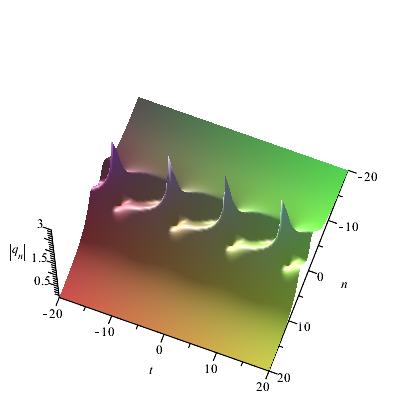}
\end{minipage}%
}%
\subfigure[]{
\begin{minipage}[t]{0.29\linewidth}
\centering
\includegraphics[width=1.8in]{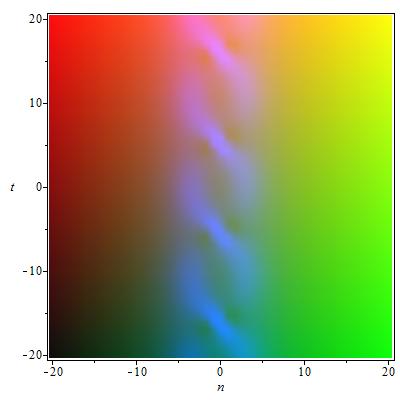}
\end{minipage}%
}%
\centering
\caption{Two-soliton solution for Eq.(\ref{ndNLS1}): (a) periodically breathing bounded two-soliton, (b)The density profiles of (a). }
\end{figure}

\subsection{Three-solitons}
Similar to one- and two- soliton solution for the reverse-space discrete DNLS equation (\ref{ndNLS1}), we obtain the
following conditions on the parameters of three-soliton solution (\ref{3soliton1}-\ref{3soliton2}) to satisfy the nonlocal reduction $r_{n}=\sigma q_{-n}^{*}$:
\begin{equation}\label{3constraints1}
\begin{aligned}
&\ l_{j}=-k_{j}^{*},\rho_{j}=\omega_{j}^{*}, e^{\alpha_{j}}=- \sigma e^{\delta_{j}} , \ j=1,2,3;\ e^{\delta_{m,j}}=e^{\alpha^{*}_{j,m}},\ m,j=1,2,3;\ K=J^{*};\\
&\ \tilde{B}_{m,j}=\tilde{A}_{m,j}^{*},B_{\mu,m,j}=A_{m,j,\mu}^{*}, \ m,j,\mu=1,2,3,m<j;\  N_{m,j,\mu,\nu}=M^{*}_{\mu,\nu,m,j},\ m,j,\mu,\nu=1,2,3,m<j, \mu<\nu.
\end{aligned}
\end{equation}
Utilizing the dispersion relations (\ref{dispersion3}) and Eqs.(\ref{3coeff1}-\ref{3coeff9}), Eq.(\ref{3constraints1}) can be reduced to the following six conditions
\begin{equation}\label{3constraints2}
l_{j}=-k_{j}^{*},\ e^{\alpha_{j}^{*}}=- \sigma e^{\delta_{j}} , \ j=1,2,3.
\end{equation}
Therefore, the 3-soliton solution of the nonlocal discrete DNLS (\ref{ndNLS1}) is given by (\ref{3soliton1}) with constraints of parameters (\ref{3constraints2}).
we choose parameters in three-soliton solution as
\begin{equation*}
\begin{aligned}
&k_{1}=0.25i, k_{2}=0.2i, k_{3}=0.8i,\delta_{1}=i,\delta_{2}=i,\delta_{3}=i,\sigma=-1,
\end{aligned}
\end{equation*}
and the corresponding three-soliton is shown in Fig.6 .

\begin{figure}[!htb]
\centering
\subfigure[]{
\begin{minipage}[t]{0.32\linewidth}
\centering
\includegraphics[width=2in]{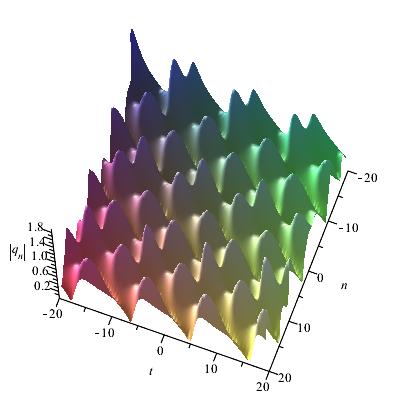}
\end{minipage}%
}%
\subfigure[]{
\begin{minipage}[t]{0.32\linewidth}
\centering
\includegraphics[width=2in]{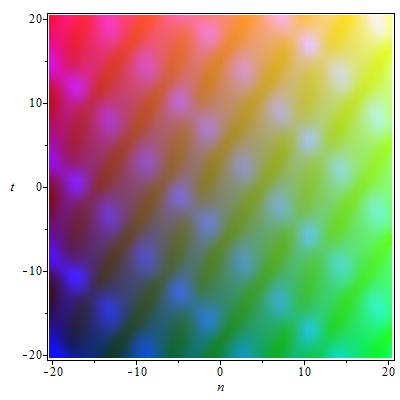}
\end{minipage}%
}%
\centering
\caption{Three-soliton solution for Eq.(\ref{ndNLS1}): (a)bounded periodic three-soliton, (b)The density profiles of (a). }
\end{figure}

\section[$r_{n}=\sigma q_{-n}(-t)$]{Soliton solutions for the reverse-space-time discrete DNLS equation (\ref{ndNLS2})}
In this section, we derive one-, two-, three-soliton solutions of the reverse-space-time discrete DNLS equation (\ref{ndNLS2}) by
finding the constraint conditions on the parameters of one-, two-, three-soliton solutions of the semi-discrete Gerdjikov-Ivanov equation (\ref{GI}) to satisfy the the reduction formula $r_{n}=\sigma q_{-n}(-t)$.

\subsection{One solitons}

From one-soliton solution (\ref{1soliton}) and reduction formula $r_{n}=\sigma q_{-n}(-t)$, we have
\begin{equation}\label{1reduction2}
-\frac{e^{l n + \rho t + \alpha }}{1+B e^{(k+l)n +(\omega+\rho) t+\delta+\alpha}}=\frac{\sigma e^{-k n-\omega t+\delta}}{1+A e^{-(k+l) n-(\omega+\rho)t+\alpha+\delta}}.
\end{equation}
By applying  the cross multiplication on (\ref{1reduction2}), we obtain
\begin{equation}
-\left(e^{ln +\rho t+\alpha}+A e^{-k n-\omega t+\delta+2\alpha}\right)=\sigma e^{-kn -\omega t+\delta}+B \sigma e^{ln +\rho t+2 \delta+\alpha},
\end{equation}
form which we derive
\begin{equation}
A e^{2 \alpha}=-\sigma, \quad B e^{2 \delta}=-\sigma,
\end{equation}
which yields $e^{\alpha}=\sqrt{-\frac{1}{\sigma A}} $ and $e^{\delta}=\sqrt{-\frac{1}{\sigma B}}.$ Therefore, one soliton solution for the reverse-space-time discrete DNLS equation (\ref{ndNLS2}) is given by
\begin{equation}\label{rsone-soliton}
q_n=\frac{e^{kn+(4i\sinh^2\frac{k}{2})t}}{\sqrt{-\sigma B}(1+\sqrt{\frac{A}{B}}e^{(k+l)n+4i(\sinh^2\frac{k}{2}-\sinh^2\frac{l}{2})t})},
\end{equation} where $k,l$ are free complex parameters. By setting $k=a+bi,c+di,\sqrt{\frac{A}{B}}=R+Ii$, the corresponding $|q_n|$ is given by

\begin{equation}\label{Modul3}
|q_n|^2=\frac{1}{|B|(e^{-2\zeta_1}+(R^2+I^2)e^{2\zeta_2}+2\sqrt{R^2+I^2}\cos(L+\gamma)e^{\zeta_2-\zeta_1})},
\end{equation}
where $\zeta_1=an-2\sinh(a)\sin(b)t,\zeta_2=cn+2\sinh(c)\sin(d)t$,$L=(b+d)n+2(\cosh(a)\cos(b)-\cosh(c)\cos(d))t,\cos(\gamma)=\frac{R}{\sqrt{R^2+I^2}},\sin(\gamma)=\frac{L}{\sqrt{R^2+I^2}}.$\\

Case \uppercase\expandafter{\romannumeral 1.} $b=d=0.$

In this case, $|q_n|$ can be written as
\begin{equation}\label{Modul31}
|q_n|^2=\frac{1}{|B|(e^{-2an}+R^2e^{2cn}+2|R|\cos(2(\cosh(a)-\cosh(c))t+\gamma)e^{(c-a)n})},
\end{equation} from which we derive that this soliton breathes and periodically collapses in $t$ with period $\frac{\pi}{\cosh(a)-\cosh(c)}$ at position $n=-\frac{\ln(\frac{e^c-1}{e^{-a}-1})}{2(a+c)}$ where the conditions $ac<0$ and $-\frac{\ln|\frac{e^c-1}{e^{-a}-1}|}{a+c}\in Z$ should be satisfied. At $n=-\frac{\ln(\frac{e^c-1}{e^{-a}-1})}{2(a+c)}$, the amplitude of the soliton changes as
\begin{equation}\label{Modul311}
|q_n|^2=\frac{1}{|B|(|R|^{\frac{2a}{a+c}}+|R|^{-\frac{2a}{a+c}}+2|R|^{\frac{2a}{a+c}}\cos(2(\cosh(a)-\cosh(c))t+\gamma))}.
\end{equation} By taking
\begin{equation*}
k=\ln\frac{2}{3},l=\ln3,\sigma=-1,\\
\end{equation*}
this soliton is illustrated in (a) of Fig.7.\\

Case \uppercase\expandafter{\romannumeral 2.} $a=c=0.$

In this case, the $|q_n|$ becomes
\begin{equation}\label{Modul32}
|q_n|^2=\frac{1}{|B|(1+R^2+I^2+2\sqrt{R^2+I^2}\cos(((b+d)n+2(\cos(b)-\cos(d))t+\gamma)}.
\end{equation}
When $R^2+I^2\neq 1,$  this soliton is bounded and periodic which is shown in (b) of Fig.7 by taking
\begin{equation*}
k=i,l=0.3i,\sigma=1.\\
\end{equation*}

Case \uppercase\expandafter{\romannumeral 3.} $a,c$ are not simultaneously zero and $b,d$ are not simultaneously zero.

In this case, this soliton moves at velocity $V=\frac{2(\sinh(a)\sin(b)-\sinh(c)\sin(d))}{a+c}$ on the line $n=Vt-\frac{1}{2(a+c)}\ln(R^2+I^2)$ where the amplitude $|q_n|$ changes as
\begin{equation*}
|q_n|^2=\frac{(R^2+I^2)^{-\frac{a}{a+c}}}{2|B|}\frac{e^{2\varrho t}}{1+\cos(\Omega t+\vartheta)},
\end{equation*}where $\varrho=aV-2\sinh(a)\sin(b),\Omega=(b+d)V+2(\cosh(a)\cos(b)-\cosh(c)\cos(d)),\vartheta=\gamma-\frac{b+d}{2(a+c)}\ln(R^2+I^2).$ When $\Omega\neq 0,$ this soliton periodically collapses with period $\frac{2\pi}{\Omega},$ and when $\varrho \neq 0,$ the amplitude of the soliton grows or decays exponentially (depending on the sign of $\varrho$) which are illustrated in (a) and (b) of Fig.8 by taking parameters as
\begin{equation*}
	\begin{aligned}
	&k=0.5-3i,l=0.6-0.5i,\sigma=1,  \\
	\end{aligned}
\end{equation*}
and
\begin{equation*}
	\begin{aligned}
	&k=0.5+3i,l=0.6+0.5i,\sigma=1, \\
	\end{aligned}
\end{equation*}
respectively.

\begin{figure}[!htb]
\centering
\subfigure[]{
\begin{minipage}[t]{0.32\linewidth}
\centering
\includegraphics[width=2in]{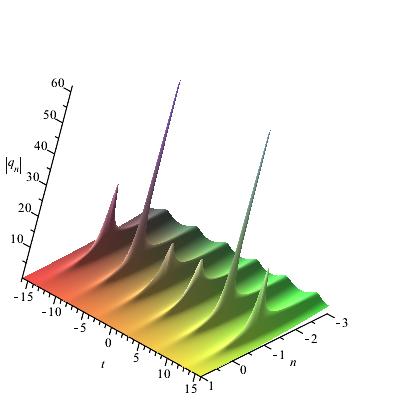}
\end{minipage}%
}%
\subfigure[]{
\begin{minipage}[t]{0.32\linewidth}
\centering
\includegraphics[width=2in]{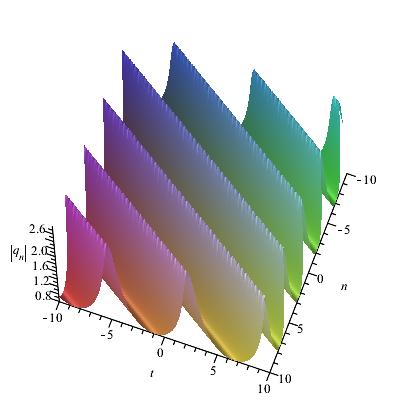}
\end{minipage}%
}%
\centering
\caption{One-soliton solution for the reverse-space-time discrete DNLS equation (\ref{ndNLS2}): (a)One-soliton breathing and periodically collapsing in time, (b) bounded periodic one-soliton. }
\end{figure}

\begin{figure}[!htb]
\centering
\subfigure[]{
\begin{minipage}[t]{0.32\linewidth}
\centering
\includegraphics[width=2in]{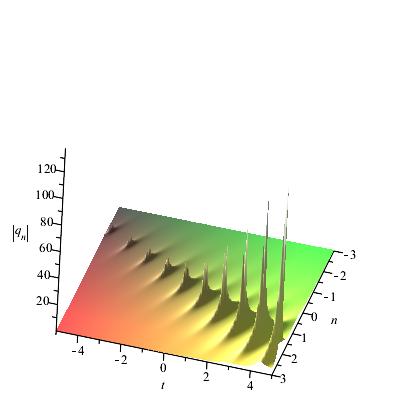}
\end{minipage}%
}%
\subfigure[]{
\begin{minipage}[t]{0.32\linewidth}
\centering
\includegraphics[width=2in]{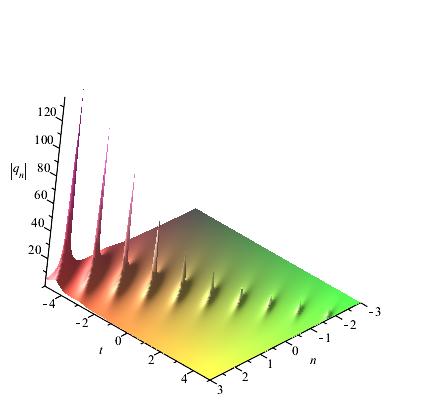}
\end{minipage}%
}%
\centering
\caption{Periodically collapsing one-soliton solution for Eq.(\ref{ndNLS2}): (a) Solution with exponentially growing amplitude, (b) Solution with exponentially decaying amplitude. }
\end{figure}

\subsection{Two-solitons}

From the two-soliton solution (\ref{2soliton1}-\ref{2soliton2}) and reduction formula $r_{n}=\sigma q_{-n}(-t)$, we have
\begin{equation}
\begin{aligned}
-\frac{e^{\eta_{1}}+e^{\eta_{2}}
	+\hat{B}_{1} e^{\xi_{1}+\eta_{1}+\eta_{2}}
	+\hat{B}_{2}e^{\xi_{2}+\eta_{1}+\eta_{2}}}{1+
	e^{\xi_{1}+\eta_{1}+\delta_{1,1}}+e^{\xi_{1}+\eta_{2}+\delta_{1,2}}
	+e^{\xi_{2}+\eta_{1}+\delta_{21}}+e^{\xi_{2}+\eta_{2}+\delta_{2,2}}
	+N e^{\xi_{1}+\xi_{2}+\eta_{1}+\eta_{2}}}
= \\
\sigma
\frac{e^{\xi_{1}^{-}}+e^{\xi_{2}^{-}}
	+\hat{A}_{1}  e^{\xi_{1}^{-}+\xi_{2}^{-}+\eta_{1}^{-}}
	+\hat{A}_{2} e^{\xi_{1}^{-}+\xi_{2}^{-}+\eta_{2}^{-}}}
{1+e^{\xi_{1}^{-}+\eta_{1}^{-}+\alpha_{1,1}}+e^{\xi_{1}^{-}+\eta_{2}^{-}+\alpha_{1,2}}
	+e^{\xi_{2}^{-}+\eta_{1}^{-}+\alpha_{2,1}}+e^{\xi_{2}^{-}+\eta_{2}^{-}+\alpha_{2,2}}
	+M e^{\xi_{1}^{-}+\xi_{2}^{-}+\eta_{1}^{-}+\eta_{2}^{-}}},
\end{aligned}
\end{equation}
where $\xi_{j}^{-}=-k_jn-\omega_j t+\delta_j,\eta_{j}^{-}=-l_in-\rho_jt+\alpha_j (j=1,2).$
Applying the cross multiplication, we get
\begin{equation}
\begin{aligned}\label{2constraints2}
&\ \hat{B}_{1} e^{2 \alpha_{j}+\alpha_{1,j}+2 \delta_{1}}+\hat{B}_{2} e^{2 \alpha_{j}+\alpha_{2,j}+2 \delta_{2}}+\sigma N \hat{A}_{j}  e^{2 \alpha_{j}+2 \delta_{1}+2 \delta_{2}}+\sigma e^{2 \delta_{1}+\delta_{1,j}}+\sigma e^{2 \delta_{2}+\delta_{2,j}}+1=0,\ j=1,2,\\
&\ \hat{A}_{1} e^{2 \delta_{j}+\delta_{j,1}+2 \alpha_{1}}+\hat{A}_{2} e^{2 \delta_{j}+\delta_{j,2}+2 \alpha_{2}}+\sigma M \hat{B}_{j}  e^{2 \delta_{j}+2 \alpha_{1}+2 \alpha_{2}}+\sigma e^{2 \alpha_{1}+\alpha_{j,1}}+\sigma e^{2 \alpha_{2}+\alpha_{j,2}}+1=0,\ j=1,2,  \\
&\ \sigma \hat{A}_{\lambda} e^{2 \delta_{\nu}+\delta_{\nu,\mu}}+e^{\alpha_{\beta,\lambda}}=0,\ \sigma\hat{B}_{\lambda} e^{2 \alpha_{\nu}+\alpha_{\mu,\nu}}+e^{\delta_{\lambda,\beta}}=0, \ \lambda,\nu\in\{1,2\};\mu\in\{1,2\}\backslash\{\lambda\}; \beta\in\{1,2\}\backslash\{\nu\},\\
&\ \sigma \hat{A}_{m}+M e^{2 \alpha_{j}}=0, \sigma \hat{B}_{m}+N e^{ 2 \delta_{j}}=0, \ 1\leq j\neq m \leq 2.
\end{aligned}
\end{equation}
Utilizing the dispersion relations (\ref{dispersion2}) and Eqs.(\ref{2coeff1}-\ref{2coeff4}), Eq.(\ref{2constraints2}) can be reduced to the following four conditions
\begin{equation}
M e^{2 \alpha_{j}}=-\sigma \hat{A}_{m},\quad  Ne^{ 2 \delta_{j}}=-\sigma \hat{B}_{m}, \quad 1\leq j \neq m \leq 2,
\end{equation}
from which we have
\begin{equation}\label{2constraint2}
e^{\alpha_{j}}= 2 \sqrt{\frac{\sinh^2\frac{k_{1}+l_{j}}{2} \sinh^2\frac{k_{2}+l_{j}}{2}}{\sigma(1-e^{l_j})\sinh^2 \frac{l_{1}-l_{2}}{2}}},\quad
e^{\delta_{j}}= 2 \sqrt{\frac{\sinh^2\frac{k_{j}+l_{1}}{2} \sinh^2\frac{k_{j}+l_{2}}{2}}{\sigma(1-e^{-k_j})\sinh^2 \frac{k_{1}-k_{2}}{2}}},\ j=1,2,
\end{equation}
where $k_{j},l_{j}(j=1,2)$ are arbitrary complex parameters. Therefore, (\ref{2soliton1}) with constraints of parameters (\ref{2constraint2}) gives two-soliton solution for the reverse-space-time discrete DNLS equation (\ref{ndNLS2}). A periodically breathing but not collapsing two-soliton solution which is asymmetric in $n$ is depicted in Fig.9 by taking the parameters as
\begin{equation*}
	\begin{aligned}
	&k_{1}=0.3,k_{2}=0.6,\l_{1}=0.6,l_{2}=0.3,\sigma=1. \\
	\end{aligned}
\end{equation*}
\begin{figure}[!htb]
\centering
\subfigure[]{
\begin{minipage}[t]{0.32\linewidth}
\centering
\includegraphics[width=2in]{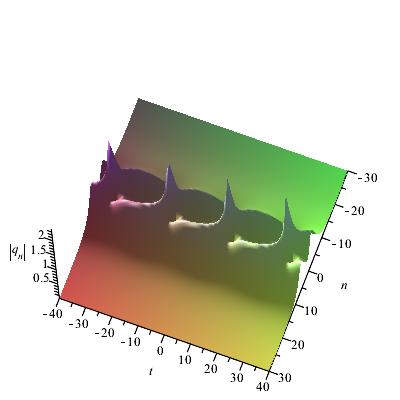}
\end{minipage}%
}%
\subfigure[]{
\begin{minipage}[t]{0.32\linewidth}
\centering
\includegraphics[width=2in]{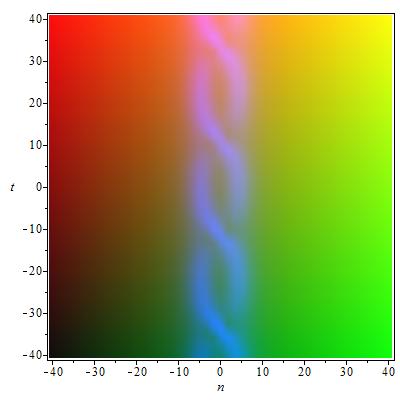}
\end{minipage}%
}%
\centering
\caption{Two-soliton solution for the reverse-space-time discrete DNLS equation (\ref{ndNLS2}): (a)Breathing 2-soliton, (b)The density profiles of (a). }
\end{figure}
 The collisions of two bounded soliton are displayed in $(a)$ and $(b)$ of Fig.10 by choosing parameters as
\begin{equation*}
	\begin{aligned}
	&k_{1}=0.3+0.5i, k_{2}=0.3-0.4i, l_{1}=0.3-0.3i, l_{2}=0.3+0.6i,\sigma=1,
	\end{aligned}
\end{equation*}	
and
\begin{equation*}
	\begin{aligned}
    &k_{1}=0.3+0.6i, k_{2}=0.3-0.6i, l_{1}=0.3-0.6i, l_{2}=0.3+0.6i,\sigma=1,
	\end{aligned}
\end{equation*}
respectively.

\begin{figure}[!htb]
\centering
\subfigure[]{
\begin{minipage}[t]{0.32\linewidth}
\centering
\includegraphics[width=2in]{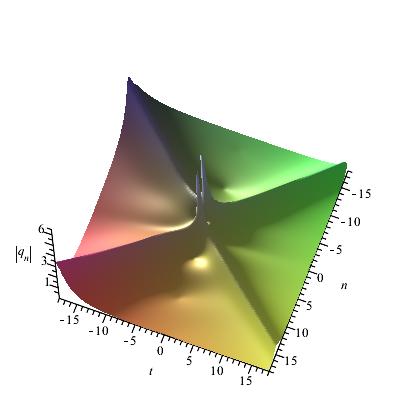}
\end{minipage}%
}%
\subfigure[]{
\begin{minipage}[t]{0.32\linewidth}
\centering
\includegraphics[width=2in]{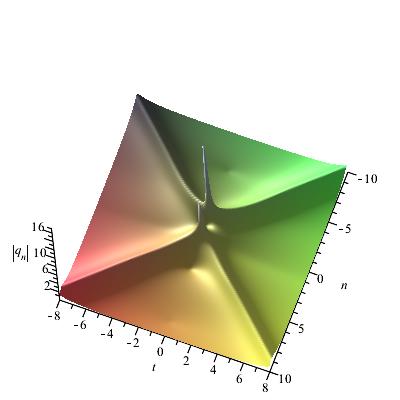}
\end{minipage}%
}%
\centering
\caption{Two-soliton solution for the reverse-space-time discrete DNLS equation (\ref{ndNLS2}): (a)Collision of two bounded soliton with exponentially decaying amplitudes, (b)Elastic collision of  two soliton.}
\end{figure}

\subsection{ Three-solitons}

By applying cross multiplication on the three-soliton solution (\ref{3soliton1}-\ref{3soliton2}) with the nonlocal reduction $r_{n}(t)=\sigma q_{-n}(-t)$, we obtain 126 constraints on parameters which are given in Appendix A.

Applying the dispersion relations (\ref{dispersion3}) and Eqs.(\ref{3coeff1}-\ref{3coeff9}), Eqs.(\ref{three1}-\ref{three6}) can be reduced to the following six constraints:
\begin{equation*}
\sigma \tilde{A}_{m,p}=-J e^{2 \alpha_{j}},\sigma \tilde{B}_{m,p}=-K e^{2 \delta_{j}},\ j\in\{1,2,3\},m,p\in\{1,2,3\}\backslash\{j\},p>m,\\
\end{equation*}
which yields
\begin{eqnarray}\label{3key}
&&e^{\alpha_{j}}= 2 \sqrt{
\frac{\sinh^{2}\frac{k_{1}+l_{j}}{2} \sinh^2\frac{k_{2}+l_{j}}{2}\sinh^2\frac{k_{3}+l_{j}}{2}}{\sigma(1-e^{l_j})\sinh^ \frac{l_{j}-l_{m}}{2}\sinh^ \frac{l_{j}-l_{p}}{2}}},j\in\{1,2,3\},m,p\in\{1,2,3\}\backslash\{j\},p>m,\label{3constraints21}\\
&&e^{\delta_{j}}= 2 \sqrt{
\frac{\sinh^{2}\frac{k_{j}+l_{1}}{2} \sinh^2\frac{k_{j}+l_{2}}{2}\sinh^2\frac{k_{j}+l_{3}}{2}}{\sigma(1-e^{-k_j})\sinh^ \frac{k_{j}-k_{m}}{2}\sinh^ \frac{k_{j}-k_{p}}{2}}},j\in\{1,2,3\},m,p\in\{1,2,3\}\backslash\{j\},p>m, \label{3constraints22}
\end{eqnarray}
where $k_{j},l_{j}(j=1,2,3)$ are arbitrary complex parameters. Therefore, Eq.(\ref{3soliton1}) with constraints on parameters (\ref{3constraints21}-\ref{3constraints22}) gives three-soliton solution for  the reverse-space-time discrete DNLS equation (\ref{ndNLS2}).
The bounded three-soliton solution which breathes periodically in $t$ is displayed in Fig.11 by taking parameters in this three-soliton solution as
\begin{equation*}
	\begin{aligned}
	&k_{1}=0.5, k_{2}=0.3,k_{3}=0.6,l_{1}=0.6,l_{2}=0.3,l_{3}=0.5,\sigma=-1.
	\end{aligned}
\end{equation*}

\begin{figure}[!htb]
\centering
\subfigure[]{
\begin{minipage}[t]{0.32\linewidth}
\centering
\includegraphics[width=2in]{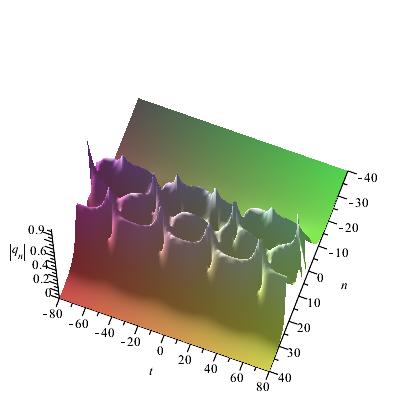}
\end{minipage}%
}%
\subfigure[]{
\begin{minipage}[t]{0.32\linewidth}
\centering
\includegraphics[width=2in]{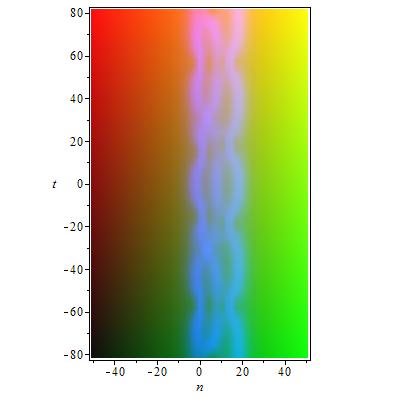}
\end{minipage}%
}%
\centering
\caption{Three-soliton solution for the reverse-space-time discrete DNLS equation (\ref{ndNLS2}): (a)Periodically breathing bounded three-soliton solution, (b)The density profiles of (a). }
\end{figure}

The interactions of three bounded solitons are displayed in Fig.12 by takeing the parameters as
\begin{equation*}
\begin{aligned}
& k_{1}=0.15+0.24i, k_{2}=0.24+0.15i,k_{3}=0.24-0.15i, l_{1}=0.15-0.24i,l_{2}=0.24-0.15i,l_{3}=0.24+0.15i,\sigma=-1.
\end{aligned}
\end{equation*}

\begin{figure}[!htb]
\centering
\subfigure[]{
\begin{minipage}[t]{0.32\linewidth}
\centering
\includegraphics[width=2in]{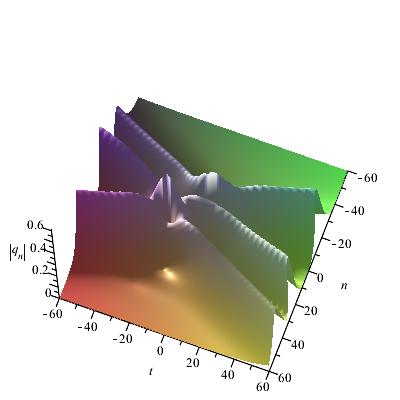}
\end{minipage}%
}%
\subfigure[]{
\begin{minipage}[t]{0.32\linewidth}
\centering
\includegraphics[width=2in]{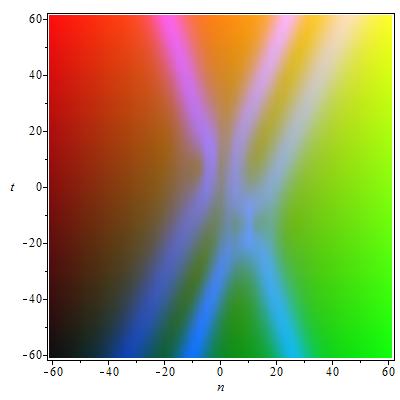}
\end{minipage}%
}%
\centering
\caption{Three-soliton solution for the reverse-space-time discrete DNLS equation (\ref{ndNLS2}): (a)Collision of bounded three soliton, (b)The density profiles of (a). }
\end{figure}

\section{Conclution and discussion}
In this paper, we proposed the reverse-space and reverse-space-time nonlocal discrete DNLS equations (\ref{ndNLS1}) and (\ref{ndNLS2}), and derived their one-, two-
and three-soliton solutions via  Hirota bilinear method and reduction approach. The dynamics of soliton solutions are discussed and rich soliton structures in the reverse-space and reverse-space-time nonlocal discrete DNLS equations are revealed. Our investigation shows that the solitons of these nonlocal equations often breathe and periodically collapse for some soliton parameters, but remain bounded for other range of parameters.

Now we investigate the continuous limit for the reverse-space nonlocal discrete DNLS equation (\ref{ndNLS1}), the reverse-space-time nonlocal discrete DNLS equation (\ref{ndNLS2}) and their one-soliton solutions. If we take
\begin{equation*}
q_n=\varepsilon Q(x,\tau),x=n \varepsilon^2,\tau=\varepsilon^4 t,
\end{equation*}
then as $\varepsilon\rightarrow 0$, Eq. (\ref{ndNLS1}) and Eq. (\ref{ndNLS2}) converge to the reverse-space and reverse-space-time nonlocal DNLS equations
\begin{equation}\label{cnDLS1}
iQ_{\tau}+Q_{xx}-2\sigma Q^2 Q^{\ast}_{x}(-x)-2Q^3{Q^{\ast}}^2(-x)=0,
\end{equation}
and
\begin{equation}\label{cnDLS2}
iQ_{\tau}+Q_{xx}-2\sigma Q^2 Q_{x}(-x,-\tau)-2Q^3Q^2(-x,-\tau)=0,
\end{equation}
respectively. Furthermore, by setting $k=\varepsilon^2 \lambda, e^{\delta}=\varepsilon e^{\beta}$ and taking limit $\varepsilon \rightarrow 0$, the first type of one-soliton solution (\ref{n1soliton1}) for the reverse-space discrete DNLS equation (\ref{ndNLS1}) converges to
\begin{equation}\label{cnDLS3}
Q(x,\tau)=\frac{1}{e^{-\lambda x-i\lambda^2\tau-\beta}+\frac{\lambda^{\ast}\sigma}{(\lambda-\lambda^{\ast})^2}e^{-\lambda^{\ast}x-i{\lambda^{\ast}}^2\tau+\beta^{\ast}}},
\end{equation}
with $\lambda,\beta$ being complex parameters, which is one type of one-soliton soluiton for the reverse-space nonlocal DNLS equation (\ref{cnDLS1}).
By setting $k=\varepsilon^2 \lambda, l=\varepsilon^2 \omega$, and taking limit $\varepsilon \rightarrow 0$, the second type of one-soliton solution (\ref{key2}) for the reverse-space discrete DNLS equation (\ref{ndNLS1}) converges to
\begin{equation}\label{cnDLS4}
Q(x,\tau)=\frac{1}{\sqrt{\frac{\sigma\lambda}{(\lambda+\omega)^2}}e^{-\lambda x-i\lambda^2\tau-bi}+\sqrt{\frac{-\sigma\omega}{(\lambda+\omega)^2}}e^{\omega x-i\omega^2\tau+di}},
\end{equation}
with $\lambda,\omega,b,d$ being real parameters, which is another type of one-soliton soluiton for the reverse-space nonlocal DNLS equation (\ref{cnDLS1}).
Setting $k=\varepsilon^2 \lambda, l=\varepsilon^2 \omega$, and taking limit $\varepsilon \rightarrow 0$, the one-soliton solution (\ref{rsone-soliton}) for the reverse-space discrete DNLS equation (\ref{ndNLS2}) converges to
\begin{equation}\label{cnDLS5}
Q(x,\tau)=\frac{1}{\sqrt{\frac{\sigma\lambda}{(\lambda+\omega)^2}}e^{-\lambda x-i\lambda^2\tau}+\sqrt{\frac{-\sigma\omega}{(\lambda+\omega)^2}}e^{\omega x-i\omega^2\tau}},
\end{equation}
with $\lambda,\omega$ being complex parameters, which is one-soliton soluiton for the reverse-space-time nonlocal DNLS equation (\ref{cnDLS2}).
The N-soliton solution expressed in terms of Grammian and Casorati determinant solutions for two types of nonlocal discrete DNLS (\ref{ndNLS1}) and (\ref{ndNLS2}) via the bilinearisation-reduction approach are under investigation.
\section*{Acknowledgements}
This work was supported by the National Natural Science Foundation of China (Grant nos. 11601247, 11965014 and 11605096).

\begin{appendices}
	
	\renewcommand{\theequation}{A.\arabic{equation}}
	\renewcommand{\thesection}{Appendix \Alph{section}}
	\section{Constrains on the parameters in three-soliton solution for the reverse-space-time discrete DNLS equation (\ref{ndNLS2})}
	Applying the cross multiplication on three-soliton solution (\ref{3soliton1}-\ref{3soliton2}) with nonlocal reduction $r_{n}(t)=\sigma q_{-n}(-t)$, we obtain the following 126 constraints on the parameters:
	\begin{eqnarray}
	&&e^{2 \delta_{\lambda}}=-\frac{ \sigma \hat{B}_{\mu,\nu}}{K}=-\frac{\sigma M_{\mu,\nu,m,p}}{ \tilde{A}_{m,p}e^{ \delta_{\lambda,j}}},\quad e^{2 \alpha_{\lambda}}=-\frac{ \sigma \tilde{A}_{\mu,\nu}}{J}=-\frac{\sigma N_{m,p,\mu,\nu}}{ \tilde{B}_{m,p}e^{\alpha_{j,\lambda}}},\nonumber\\
	&&\qquad\qquad\qquad\qquad \lambda,j \in \{1,2,3\};\mu,\nu\in \{1,2,3\}\backslash \{\lambda\},\nu>\mu;m,p\in \{1,2,3\}\backslash \{j\},p>m,\label{three1}\\
	&&e^{2 \delta_{\lambda}}=-\frac{\sigma B_{\mu,m,p} e^{\alpha_{\nu,j}}}{A_{\nu,\lambda,j} N_{\lambda,\mu,m,p}}, \quad e^{2 \alpha_{\lambda}}=-\frac{\sigma A_{m,p,\mu} e^{\delta_{j,\nu}}}{B_{j,\nu,\lambda} M_{m,p,\lambda,\mu}},\nonumber\\
	&&\qquad\qquad\qquad\qquad \lambda,j \in \{1,2,3\};\mu,\nu\in \{1,2,3\}\backslash \{\lambda\};m,p\in \{1,2,3\}\backslash \{j\},p>m,\label{three2}\\ \nonumber\\
&&e^{2\delta_{1}}e^{2\delta_{2}}A_{{1,2,\lambda}}N_{{1,2,\mu,\nu}}
	+e^{2\delta_{1}}e^{2\delta_{3}}A_{{1,3,\lambda}}N_{{1,3,\mu,\nu}}
	+e^{2\delta_{2}}e^{2\delta_{3}}A_{{2,3,\lambda}}N_{{2,3,\mu,\nu}}+\sigma\,e^{2\delta_{1}}e^{\alpha_{1,\lambda}}B_{{1,\mu,\nu}}\nonumber\\
	&&\quad+\sigma\, e^{2\delta_{2}}e^{\alpha_{2,\lambda}}B_{{2,\mu,\nu}}+\sigma\, e^{2\delta_{3}}e^{\alpha_{3,\lambda}}B_{{3,\mu,\nu}}=0, \nonumber\\
	&&e^{2\alpha_{1}}e^{2\alpha_{2}}B_{{\lambda,1,2}}M_{{\mu,\nu,1,2}}
	+e^{2\alpha_{1}}e^{2\alpha_{3}}B_{{\lambda,1,3}}M_{{\mu,\nu,1,3}}
	+e^{2\alpha_{2}}e^{2\alpha_{3}}B_{{\lambda,2,3}}M_{{\mu,\nu,2,3}}
	+\sigma\,e^{2\alpha_{1}}e^{\delta_{\lambda,1}}A_{{\mu,\nu,1}}\nonumber \\
	&&\quad+\sigma\,e^{2\alpha_{2}}e^{\delta_{\lambda,2}}A_{{\mu,\nu,2}}+\sigma\,e^{2\alpha_{3}}e^{\delta_{\lambda,3}}A_{{\mu,\nu,3}}=0,\qquad \lambda\in\{1,2,3\};\mu,\nu\in\{1,2,3\}\backslash\{\lambda\},\nu>\mu,\label{three3}\\ \nonumber\\
	&&B_{{\lambda,m,p}}
	+\sigma\,e^{2\delta_{\nu}}N_{{\lambda,\nu,m,p}}+\sigma\,e^{2\delta_{\mu}}N_{{\lambda,\mu,m,p}}
	+e^{2\delta_{\mu}}e^{2\alpha_{j}}e^{\alpha_{\mu,j}}\tilde{B}_{{\lambda,\mu}}
	+e^{2\delta_{\nu}}e^{2\alpha_{j}}e^{\alpha_{\nu,j}}  \tilde{B}_{{\lambda,\nu}}+\sigma\,e^{2\delta_{\mu}}e^{2\delta_{\nu}}e^{2\alpha_{j}}A_{{\mu,\nu,j}}K\nonumber\\
	&&\quad=0,\nonumber\\
	&&A_{{m,p,\lambda}}
	+\sigma\,e^{2\alpha_{\nu}}M_{{m,p,\lambda,\nu}}+\sigma\,e^{2\alpha_{\mu}}M_{{m,p,\lambda,\mu}}
	+e^{2\alpha_{\mu}}e^{2\delta_{j}}e^{\delta_{j,\mu}}\tilde{A}_{{\lambda,\mu}}
	+e^{2\alpha_{\nu}}e^{2\delta_{j}}e^{\delta_{j,\nu}} \tilde{A}_{{\lambda,\nu}}+\sigma\,e^{2\alpha_{\mu}}e^{2\alpha_{\nu}}e^{2\delta_{j}}B_{{j,\mu,\nu}}J\nonumber\\
	&&\quad=0,  \qquad \lambda,j\in\{1,2,3\};\mu,\nu\in\{1,2,3\}\backslash\{\lambda\},\nu>\mu;m,p\in\{1,2,3\}\backslash\{j\},p>m,\label{three4}\\ \nonumber\\
	&&\sigma\,e^{2\delta_{m}}e^{2\delta_{p}}e^{2\alpha_{\kappa}}\tilde{A}_{{\mu,\nu}}N_{{m,p,\kappa,\lambda}}
	+\sigma\,e^{2\delta_{m}}e^{\delta_{m,\lambda}}A_{{j,m,\beta}}
	+\sigma\,e^{2\delta_{p}}e^{\delta_{p,\lambda}}A_{{j,p,\beta}}
	+e^{2\delta_{m}}e^{2\alpha_{\kappa}}B_{{m,\kappa,\lambda}}M_{{j,m,\mu,\nu}}\nonumber \\
	&&\quad+e^{2\delta_{p}}e^{2\alpha_{\kappa}} B_{{p,\kappa,\lambda}}M_{{j,p,\mu,\nu}}+e^{\alpha_{j,\beta}}=0,\nonumber\\
	&&\sigma\,e^{2\alpha_{m}}e^{2\alpha_{p}}e^{2\delta_{\kappa}}\tilde{B}_{{\mu,\nu}}M_{{\kappa,\lambda,m,p}}
	+\sigma\,e^{2\alpha_{m}}e^{\alpha_{\lambda,m}}B_{{\beta,j,m}}
	+\sigma\,e^{2\alpha_{p}}e^{\alpha_{\lambda,p}}B_{{\beta,j,p}}
	+e^{2\alpha_{m}}e^{2\delta_{\kappa}}A_{{\kappa,\lambda,m}}N_{{\mu,\nu,j,m}}\nonumber \\
	&&\quad+e^{2\alpha_{p}}e^{2\delta_{\kappa}} A_{{\kappa,\lambda,p}}M_{{\mu,\nu,j,p}}+e^{\delta_{\beta,j}}=0,\quad \lambda,j\in\{1,2,3\};\mu,\nu\in\{1,2,3\}\backslash\{\lambda\},\nu>\mu;\beta\in\{1,2,3\}\backslash\{\lambda\};\nonumber\\
	&& \qquad\kappa\in\{1,2,3\}\backslash\{\beta,\lambda\};m,p\in\{1,2,3\}\backslash\{j\},p>m,\label{three5}\\ \nonumber
	\end{eqnarray}
	\begin{eqnarray}
	&&\sigma\,e^{2\delta_{1}}e^{2\delta_{2}}e^{2\alpha_{\mu}}A_{{1,2,\mu}}N_{{1,2,\mu,\lambda}}
	+\sigma\,e^{2\delta_{1}}e^{2\delta_{3}}e^{2\alpha_{\mu}}A_{{1,3,\mu}}N_{{1,3,\mu,\lambda}}
	+\sigma\,e^{2\delta_{1}}e^{2\delta_{2}}e^{2\alpha_{\nu}}A_{{1,2,\nu}}N_{{1,2,\nu,\lambda}}\nonumber\\
	&&+\sigma\,e^{2\delta_{1}}e^{2\delta_{3}}e^{2\alpha_{\nu}}A_{{1,3,\nu}}N_{{1,3,\nu,\lambda}}
	+\sigma\,e^{2\delta_{2}}e^{2\delta_{3}}e^{2\alpha_{\mu}}A_{{2,3,\mu}}N_{{2,3,\mu,\lambda}}
	+\sigma\, e^{2\delta_{2}}e^{2\delta_{3}}e^{2\alpha_{\nu}} A_{{2,3,\nu}}N_{{2,3,\nu,\lambda}}\nonumber\\
	&&+e^{2\delta_{1}}e^{2\alpha_{\mu}}e^{\alpha_{1,\mu}} B_{{1,\mu,\lambda}}
	+e^{2\delta_{1}}e^{2\alpha_{\nu}}e^{\alpha_{1,\nu}} B_{{1,\nu,\lambda}}
	+e^{2\delta_{2}}e^{2\alpha_{\mu}}e^{\alpha_{2,\mu}}B_{{2,\mu,\lambda}}
	+e^{2\delta_{2}}e^{2\alpha_{\nu}}e^{\alpha_{2,\nu}}B_{{2,\nu,\lambda}}+\nonumber\\
	&&e^{2\delta_{3}}e^{2\alpha_{\mu}}e^{\alpha_{3,\mu}}B_{{3,\mu,\lambda}}
	+e^{2\delta_{3}}e^{2\alpha_{\nu}}e^{\alpha_{3,\nu}}B_{{3,\nu,\lambda}}+e^{2\delta_{1}}e^{2\delta_{2}}e^{2\alpha_{\mu}}e^{2\alpha_{\nu}}\tilde{B}_{{1,2}}M_{{1,2,\mu,\nu}}
	+e^{2\delta_{1}}e^{2\delta_{3}}e^{2\alpha_{\mu}}e^{2\alpha_{\nu}}\tilde{B}_{{1,3}}M_{{1,3,\mu,\nu}}\nonumber\\
	&&+e^{2\delta_{2}}e^{2\delta_{3}}e^{2\alpha_{\mu}}e^{2\alpha_{\nu}}\tilde{B}_{{2,3}}M_{{2,3,\mu,\nu}}
	+\sigma\,e^{2\delta_{1}}e^{\delta_{1,\lambda}}+\sigma\,e^{2\delta_{2}}e^{\delta_{2,\lambda}}+
	\sigma\,e^{2\delta_{3}}e^{\delta_{3,\lambda}}+\sigma\,e^{2\delta_{1}}e^{2\delta_{2}}e^{2\delta_{3}}e^{2\alpha_{\mu}}e^{2\alpha_{\nu}}\tilde{A}_{{\mu,\nu}}K \nonumber\\
	&&\quad+1=0,\nonumber\\
	&&\sigma\,e^{2\alpha_{1}}e^{2\alpha_{2}}e^{2\delta_{\mu}}B_{{\mu,1,2}}M_{{\mu,\lambda,1,2}}
	+\sigma\,e^{2\alpha_{1}}e^{2\alpha_{3}}e^{2\delta_{\mu}}B_{{\mu,1,3}}M_{{\mu,\lambda,1,3}}
	+\sigma\,e^{2\alpha_{1}}e^{2\alpha_{2}}e^{2\delta_{\nu}}B_{{\nu,1,2}}M_{{\nu,\lambda,1,2}}\nonumber\\ &&+\sigma\,e^{2\alpha_{1}}e^{2\alpha_{3}}e^{2\delta_{\nu}}B_{{\nu,1,3}}M_{{\nu,\lambda,1,3}}+\sigma\,e^{2\alpha_{2}}e^{2\alpha_{3}}e^{2\delta_{\mu}}B_{{\mu,2,3}}M_{{\mu,\lambda,2,3}}
	+\sigma\, e^{2\alpha_{2}}e^{2\alpha_{3}}e^{2\delta_{\nu}} B_{{\nu,2,3}}M_{{\nu,\lambda,2,3}}\nonumber\\
	&&+e^{2\alpha_{1}}e^{2\delta_{\mu}}e^{\delta_{\mu,1}}A_{{\mu,\lambda,1}}
	+e^{2\alpha_{1}}e^{2\delta_{\nu}}e^{\delta_{\nu,1}} A_{{\nu,\lambda,1}}
	+e^{2\alpha_{2}}e^{2\delta_{\mu}}e^{\delta_{\mu,2}}A_{{\mu,\lambda,2}}
	+e^{2\alpha_{2}}e^{2\delta_{\nu}}e^{\delta_{\nu,2}}A_{{\nu,\lambda,2}}+\nonumber\\
	&&e^{2\alpha_{3}}e^{2\delta_{\mu}}e^{\delta_{\mu,3}}A_{{\mu,\lambda,3}}
	+e^{2\alpha_{3}}e^{2\delta_{\nu}}e^{\delta_{\nu,3}}A_{{\nu,\lambda,3}}+e^{2\alpha_{1}}e^{2\alpha_{2}}e^{2\delta_{\mu}}e^{2\delta_{\nu}}\tilde{A}_{{1,2}}N_{{\mu,\nu,1,2}}
	+e^{2\alpha_{1}}e^{2\alpha_{3}}e^{2\delta_{\mu}}e^{2\delta_{\nu}}\tilde{A}_{{1,3}}N_{{\mu,\nu,1,3}}\nonumber\\
	&&+e^{2\alpha_{2}}e^{2\alpha_{3}}e^{2\delta_{\mu}}e^{2\delta_{\nu}}\tilde{A}_{{2,3}}N_{{\mu,\nu,2,3}}
	+\sigma\,e^{2\alpha_{1}}e^{\alpha_{\lambda,1}}+\sigma\,e^{2\alpha_{2}}e^{\alpha_{\lambda,2}}
	+\sigma\,e^{2\alpha_{3}}e^{\alpha_{\lambda,3}}+\sigma\,e^{2\alpha_{1}}e^{2\alpha_{2}}e^{2\alpha_{3}}e^{2\delta_{\mu}}e^{2\delta_{\nu}}\tilde{B}_{{\mu,\nu}}J\nonumber\\
	&&\quad +1=0,\qquad\qquad\qquad\qquad\qquad\qquad\quad  \lambda\in\{1,2,3\};\mu,\nu\in\{1,2,3\}\backslash\{\lambda\},\nu>\mu.\label{three6}
	\end{eqnarray}
\end{appendices}

\end{document}